\documentclass[prb,notitlepage,groupedaddress,twocolumn,showpacs]{revtex4-1}
\usepackage{amsmath,amssymb,graphicx,xcolor,units,hyperref}

\newcommand{\dd}[1]{\mathrm{d}#1\,}
\newcommand{\avg}[1]{\langle{#1}\rangle}

\renewcommand{\Im}{\mathop{\mathrm{Im}}}

\newcommand{\sgn}{\mathop{\mathrm{sgn}}}

\newcommand{\sinc}{\mathop{\mathrm{sinc}}}
\newcommand{\sinhc}{\mathop{\mathrm{sinhc}}}

\newcommand{\bra}[1]{\langle{#1}\rvert}
\newcommand{\ket}[1]{\lvert{#1}\rangle}

\begin{document}

\title{Signatures of Rashba spin-orbit interaction in the superconducting
  proximity effect in helical Luttinger liquids}
\author{Pauli Virtanen}
\affiliation{Institute for Theoretical Physics and Astrophysics,
  University of W\"urzburg, D-97074 W\"urzburg, Germany}
\author{Patrik Recher}
\affiliation{Institute for Theoretical Physics and Astrophysics,
  University of W\"urzburg, D-97074 W\"urzburg, Germany}
\affiliation{Institute for Mathematical Physics, TU Braunschweig,
  38106 Braunschweig, Germany}
\date{\today}


\begin{abstract}
  We consider the superconducting proximity effect in a helical
  Luttinger liquid at the edge of a 2D topological insulator, and
  derive the low-energy Hamiltonian for an edge state tunnel-coupled
  to a $s$-wave superconductor. In addition to correlations between
  the left and right moving modes, the coupling can induce them inside
  a single mode, as the spin axis of the edge modes is not necessarily
  constant. This can be induced controllably in HgTe/CdTe quantum
  wells via the Rashba spin-orbit coupling, and is a consequence of
  the 2D nature of the edge state wave function. The distinction of
  these two features in the proximity effect is also vital for the use
  of such helical modes in order to split Cooper-pairs. We discuss the
  consequent transport signatures, and point out a long-ranged feature
  in a dc conductance measurement that can be used to distinguish the
  two types of correlations present and to determine the magnitude of
  the Rashba interaction.
\end{abstract}

\maketitle

\section{Introduction}

The helical edge states of a 2D topological insulator (TI) consist of
a Kramers pair of right- and left-moving electron modes of opposite
spin situated inside the bulk gap \cite{kane2005-qsh,
  bernevig2006-qsh, wu2006-hla, xu2006-soq}, and they have so far been
observed in HgTe/CdTe quantum wells (HgTe-QW)
\cite{bernevig2006-qsh,konig2007-qsh,roth2009-nti}.  In 3D topological
insulators, the edge states cover the surface of the material and
consist of a single-valley Dirac cone with spin-momentum locking,
which leads to unique electromagnetic properties and quantum
interference effects \cite{qi2011-tia,*hasan2010-cti}. In both 2D-TI
and 3D-TI the coupling of spin and orbital motion can lead to
interesting effects when combined with
superconductivity. Superconducting correlations induced by the
proximity of a singlet $s$-wave superconductor can inside the TI
obtain a $p$-wave character, which can be used to engineer Majorana
bound states.  \cite{fu2008-spe,linder2010-ust,stanescu2010-pea} A
somewhat similar induction of non-conventional correlations has also
been proposed to occur in other semiconductor systems in the combined
presence of the spin-orbit interaction and
superconductivity. \cite{sau2010-gnp,*sau2010-rom,*alicea2010-mfi,*linder2010-mfm}

When the edge state of a 2D-TI is coupled to a singlet superconductor,
the transfer of electrons between the systems can, first of all,
induce singlet-type proximity correlations between electrons in the
right and left moving modes (the $+-$ channel).  \cite{fu2008-spe}
This already leads to several effects of interest. For instance, the
helicity of the electron liquid lifts the spin degeneracy and enables
Majorana states, \cite{fu2009-jca} causes Cooper pairs to split,
\cite{sato2010-cii} and affects transport properties.
\cite{adroguer2010-phe} Tight-binding calculations studying the pair
amplitude have also been made \cite{black-schaffer2011-ssp}. There is,
however, also a possibility of inducing correlations only within the
right-moving (or the left-moving) channel at a nonzero total momentum
(the $++$ and $--$ channels). Such a channel is not forbidden by symmetries
in the problem: due to the spin-orbit coupling, the spin axis of the
edge state is not necessarily constant, so that the electrons forming
a Cooper pair singlet can both enter the same mode on the TI edge,
even when spin is conserved in the tunneling process and time-reversal
symmetry is present. In HgTe-QW, a non-constant spin axis can be
induced externally by the Rashba spin-orbit coupling that breaks
inversion symmetry. \cite{roth2009-nti} Momentum conservation is
required to be broken, but this can occur e.g. due to inhomogeneity or
a finite size of a tunneling contact.  Moreover, unlike in metals, in
2D-TI the momentum non-conservation can in principle be made
arbitrarily small by tuning the Fermi level near the Dirac point
($k=0$).

A straightforward way to probe the existence of superconducting
correlations is to observe the Josephson effect or other interference
effects that can be modulated with superconducting phase
differences. The Josephson effect has been studied previously in
various one-dimensional Luttinger liquid systems.
\cite{fisher1994-cti,fazio1996-daa,fazio1995-jct} The finite-momentum
channel has, however, received limited attention,
\cite{pugnetti2007-dje} and is usually negligible. As shown below,
certain experiments with superconducting contacts attached to the
helical edge states can nevertheless probe such microscopic aspects of
the tunneling, including the role of the Rashba interaction.

Here, we first derive a low-energy Hamiltonian describing the
superconducting proximity effect in the edge states of a 2D TI coupled
to a conventional superconductor by tunnel contacts.  We use it to
find the signatures of both types of tunneling events in a transport
experiment. Because of the reduced number of propagating modes in the
helical liquid, correlations within the same channel occur at a finite
momentum and, as in chiral liquids, \cite{fisher1994-cti} are affected
by the exclusion principle. It turns out that although this component
of the proximity effect gives a negligible correction to the dc
Josephson effect, in the NS tunneling conductance [see
Fig.~\ref{fig:setup}(c)] it manifests as a long-ranged interference
effect, oscillating as a function of the superconducting phase
difference, and unlike the $+-$ part, is not exponentially
suppressed at length scales longer than the thermal wavelength. The
ratio of the contributions of the two possible channels scales as
$\delta G_{++}/\delta G_{+-}\propto{}(z_0/\hbar
v_F)^2(k_BT/M)^2e^{2\pi Td/\hbar v_F}$ (in the noninteracting case),
where $z_0$ characterizes the strength of Rashba interaction, $v_F$ is
the Fermi velocity of the edge channels, $M$ is the energy gap of the
TI, and $d$ the distance between two superconducting contacts forming
the interferometry setup. The amplitude of the effect is proportional
to the amount of spin rotation achieved by Rashba interaction, and the
quadratic temperature dependence is due to the exclusion principle.
We also discuss how $e$-$e$ interactions modify this result.

This paper is organized as follows.  In Section \ref{sec:model}, we
introduce the model for the helical Luttinger liquid (HLL), the
coupling to the superconductors, and the electronic structure of
HgTe-QWs. Section \ref{sec:low-energy} discusses the effective
low-energy Hamiltonian, and Section \ref{sec:transport} transport
signatures in the dc and ac Josephson effects and the NS
conductance. Section \ref{sec:discussion} concludes the manuscript
with a discussion on the results and remarks on experimental
realizability.

\section{Model}
\label{sec:model}

\begin{figure}
  \includegraphics{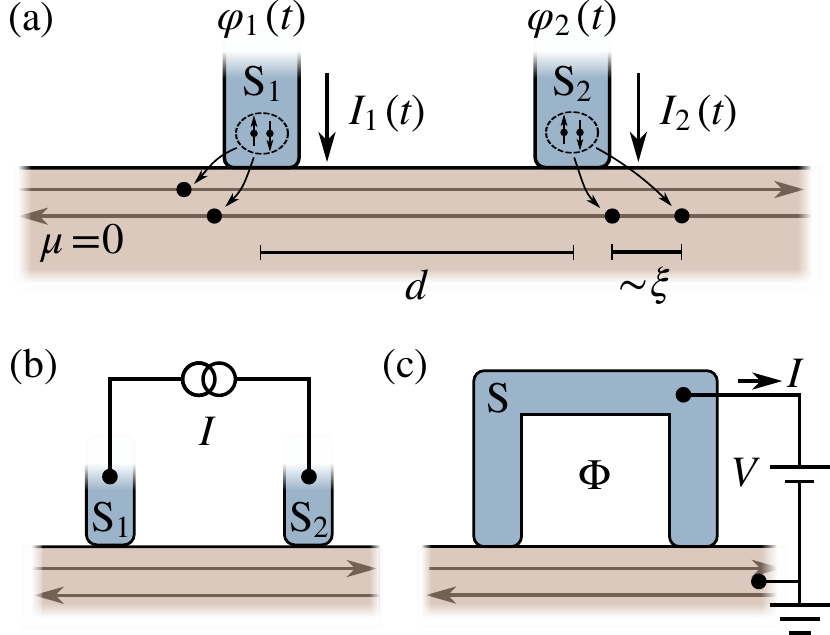}
  \caption{\label{fig:setup}
    (a)
    The setup considered: a 2D topological
    insulator, whose edge state is coupled to two superconductors via
    tunnel contacts.  In response to phase $\varphi$ or voltage $V$
    differences between the superconductors, Josephson currents can
    flow via the edge channel, or one can probe the NS transmission by
    injecting current from the superconductors to the edge channel.
    If the spin axis of the edge state is not constant spatially and as
    a function of energy, a Cooper pair singlet
    can enter the edge state in two possible ways: either electrons
    enter modes propagating to opposite directions (left) or the same
    direction (right). In the latter case, exclusion principle requires
    temporal (or spatial) separation of the two, which can be of the order
    of the superconducting coherence length $\xi=\hbar v_F/\Delta$ still
    preserving the correlation.
    (b) Configuration for the measurement of the Josephson current.
    (c) Configuration for the measurement of interference in the NS conductance.
  }
\end{figure}

We consider the setup depicted in Fig.~\ref{fig:setup}.  The edge
states of a 2D-TI are coupled to two superconducting terminals via two
tunnel junctions.  Below, we in general assume that the distance $d$
between the contacts is longer than the superconducting coherence
length $\xi$.

The left- and right moving edge states $\ket{+,x}$ and $\ket{-,x}$
have a linear dispersion, and are described by the bosonized
Hamiltonian  \cite{wu2006-hla}
\begin{equation}
  \label{eq:bosonized-hamiltonian}
    H_0 = \frac{1}{2}\int_{-\infty}^\infty\dd{x}u
    [g^{-1}(\partial_x\vartheta)^2 + g(\partial_x\phi)^2]
\end{equation}
where the Fermi field operator is $\psi_\alpha(x)=(2\pi
a_0)^{-1/2}U_\alpha e^{i\alpha k_F x}e^{i\phi_{\alpha}(x)}= (2\pi
a_0)^{-1/2}U_\alpha e^{i\alpha[k_Fx
  +\sqrt{\pi}\vartheta(x)]+i\sqrt{\pi}\phi(x)}$, the standard boson
fields $\vartheta(x)$, $\phi(x)$ satisfy
$[\phi(x),\vartheta(x')]=(i/2)\sgn(x-x')$, and $U_\pm$ are the Klein
factors.  $u=v_F/g$ is the renormalized Fermi velocity. Here and
below, we let $\hbar=k_B=e=1$, unless otherwise mentioned.  The
parameter $a_0$ is the short-distance cutoff.  In the noninteracting
case, the Luttinger interaction parameter $g=1$, and with repulsive
electron-electron interactions one has $g<1$.

The coupling to the superconductors is modeled with a tunneling
Hamiltonian
\begin{align}
  \label{eq:tunnel-hamiltonian}
  H_T =
  \sum_{\alpha=\pm, \sigma'=\uparrow,\downarrow}
  \int\dd{x}\dd{^3 r'} t_{\alpha\sigma'}(x,\vec{r}')
  \psi^\dagger_{\alpha}(x)\psi_{S\sigma'}(\vec{r}')
  +
  \mathrm{h.c.}
  \,,
\end{align}
where the tunneling amplitude $t_{\alpha\sigma'}(x,\vec{r}')$ describes the
tunneling from the state $\ket{\sigma',\vec{r}'}$ in the
superconductor to state $\ket{\alpha,x}$ in the edge mode.  For what
follows, it is useful to introduce also the corresponding one-particle
operator $\hat{h}_T$, in terms of which,
$
  t_{\alpha\sigma'}(x,\vec{r}')
  \equiv{}
  \bra{\alpha,x}\hat{h}_T\ket{\sigma',\vec{r}'}
$
  \,.
The momentum $k$ along the edge is a good quantum number for
straight TI edges, and we define the state $\ket{\alpha,x}$ in the
momentum representation:
$\ket{\alpha,x}=\sum_{k}e^{-ikx}\ket{\alpha,k}$, where
$\ket{\alpha,k}$ is the edge eigenstate with momentum $k$ and
propagation direction $\alpha=\pm$.

We assume that the Hamiltonian is time-reversal symmetric, which
implies that the tunneling operator in general satisfies
$\mathfrak{T}\hat{h}_T\mathfrak{T}^{-1}=\hat{h}_T$. Here, we choose
the phases of the wave functions so that the time reversal operations
read
$\mathfrak{T}\ket{\sigma',\vec{r}'}=\sigma'\ket{-\sigma',\vec{r}'}$
and $\mathfrak{T}\ket{\alpha,k}=\alpha\ket{-\alpha,-k}$.  We also
assume that the tunneling is spin-conserving, that is, written in
terms of real electron spin states in the TI and the superconductor,
we have $\bra{\sigma,\vec{r}}\hat{h}_T\ket{-\sigma,\vec{r}'}=0$.

To describe tunneling to HgTe-QWs, we need some knowledge of the
structure of the edge states.  This can be obtained from the four-band
model used in Ref.~\onlinecite{bernevig2006-qsh}.  In this approach,
the low-energy properties of the TI are described using a 2D envelope
function in the basis of four states
$\{\ket{E1+},\ket{H1+},\ket{E1-},\ket{H1-}\}$ localized in the quantum
well. \cite{bernevig2006-qsh,novik2005-bso} The edge states at the
boundaries of the TI can be solved within this four-band model;
\cite{zhou2008-fse} for which we give a full analytical solution in
Appendix~\ref{sec:edge-states}.

We assume the terminals are conventional spin-singlet superconductors.
As usual, \cite{abrikosov1975-moq} they are characterized by
the correlation function
$F(\vec{r}_1,\sigma_1,\tau_1;\vec{r}_2,\sigma_2,\tau_2)\equiv\avg{T[\psi_{\sigma_1}(\vec{r}_1,\tau_1)\psi_{\sigma_2}(\vec{r}_2,\tau_2)]}_0$
that has a singlet symmetry
$F(\vec{r}_1,\sigma_1,\tau_1;\vec{r}_2,\sigma_2,\tau_2)=\sigma_1\delta_{\sigma_1,-\sigma_2}F(\vec{r}_1,\tau_1;\vec{r}_2,\tau_2)$.
In the bulk, the correlation function obtains its equilibrium BCS
form, which in imaginary time can be written as
\begin{align}
  \label{eq:F-function}
  F(\vec{r}_1,\vec{r}_2;\omega)
  =
  \int\frac{\dd{^3k}}{(2\pi)^3} e^{-i(\vec{r}_1-\vec{r}_2)\cdot \vec{k}}
  \frac{\Delta}{\omega^2 + \xi_k^2 + |\Delta|^2}
  \,,
\end{align}
with $\xi_k = k^2/(2m) - \mu$ the dispersion relation and $\Delta$ the gap
of the superconductor.

\section{Effective Hamiltonian}
\label{sec:low-energy}

Integrating out the superconductors using perturbative renormalization
group theory (RG) and
considering only energies $|E|\ll|\Delta|$ reduces the Hamiltonian
$H_0+H_T$ of the total system to one concerning only the
one-dimensional edge states:
\begin{align}
  \label{eq:effective-hamiltonian}
  H
  &=
  H_0
  +
  \int\dd{x}[
  \Gamma_{+-}(x) \psi_+(x)\psi_-(x)
  \\\notag
  &\qquad
  +
  \Gamma_{++}(x)\psi_+(x)\psi_+(x+a)
  \\\notag
  &\qquad
  +
  \Gamma_{--}(x)\psi_-(x)\psi_-(x+a)
  +
  \mathrm{h.c.}
  ]
  \,.
\end{align}
Here, $\Gamma_{\alpha\beta}$ describe the coupling to the
superconductor, and $a=\hbar v_F/\Delta$ is the new
short-distance cutoff in the theory. Details of the derivation are
discussed in Appendix~\ref{sec:low-energy-hamiltonian}.

The coupling factors in the noninteracting case ($g=1$) are
given by the expressions (see Appendix~\ref{sec:low-energy-hamiltonian}
for general discussion):
\begin{gather}
  \label{eq:gamma-plus-plus}
  \Gamma_{++}(x)
  =
  \frac{\pi}{2}
  \int
  \dd{^3r_1'}\dd{^3r_2'}
  \sum_{K}
  e^{iKx}
  F(r_1',r_2';0)
  \\\notag\times
  \Delta v_F^{-1}\partial_k
  P_{++}(\frac{K}{2}+k,r_1';\frac{K}{2}-k,r_2')^*
  \rvert_{k=0}
  \,,
\end{gather}
where $F$ is given in Eq.~\eqref{eq:F-function}, and
\begin{gather}
  \label{eq:gamma-plus-minus-expression}
  \Gamma_{+-}(x)
  =
  \pi
  \int
  \dd{^3r_1'}\dd{^3r_2'}
  \sum_{K}
  e^{iKx}
  F(r_1',r_2';0)
  \\\notag
  \times P_{+-}(\frac{K}{2}-k_F,r_1';\frac{K}{2}+k_F,r_2')^*
  \,,
\end{gather}
with $\Gamma_{--}=\Gamma_{++}^*$ in the presence of time reversal
symmetry.  The main contributions should arise around $K=-2k_F$ for
$\Gamma_{++}$ (due to the $2k_F$ oscillations in the Fermi operators),
and around $K=0$ for $\Gamma_{+-}$.

\begin{figure}
  \includegraphics{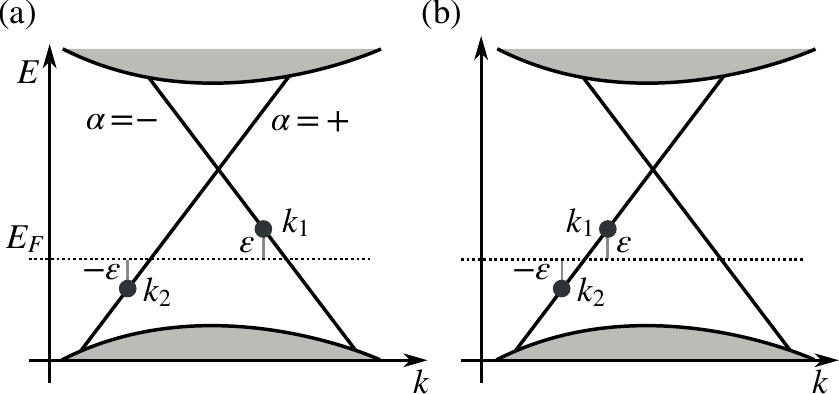}
  \caption{\label{fig:pairing-schematics} 
    Schematic depiction of the edge mode spectrum and the states
    participating in 2-particle tunneling from the Fermi level in the
    superconductor to states in the TI. 
    (a) In the $+-$ channel, 
    the momenta corresponding to energy $\varepsilon$ are
    $k_1=k_F-\varepsilon/\hbar v_F$,
    $k_2=-k_F-\varepsilon/\hbar v_F$.
    (b) In the $++$ channel, one has
    $k_1=-k_F+\varepsilon/\hbar v_F$,
    $k_2=-k_F-\varepsilon/\hbar v_F$.
    Singlet pair tunneling into the $++$ channel can occur 
    if the electron spin axis
    is different for the states at $k_1$ and $k_2$.
  }
\end{figure}

The coupling is proportional to the factor
\begin{align}
  \label{eq:M-tunneling-element}
  P_{\alpha_1\alpha_2}(k_1,\vec{r}_1';k_2,\vec{r}_2')
  &\equiv
  [
  t_{\alpha_1\downarrow}(k_1,\vec{r}_1')t_{\alpha_2\uparrow}(k_2,\vec{r}_2')
  \\
  \notag
  &
  -
  t_{\alpha_1\uparrow}(k_1,\vec{r}_1')t_{\alpha_2\downarrow}(k_2,\vec{r}_2')
  ]
  +
  [
  \vec{r}_1'\leftrightarrow{}\vec{r}_2'
  ]
  \,,
\end{align}
which describes two-particle tunneling of a singlet, from two points
$\vec{r}_1'$ and $\vec{r}_2'$ in the superconductor, to momentum states
$\ket{\alpha_1,-k_1}$, $\ket{\alpha_2,-k_2}$ in the TI edge modes
(cf. Fig.~\ref{fig:pairing-schematics}).  Here,
$t_{\alpha\sigma'}(k,\vec{r}') =
\int\dd{x}e^{ikx}t_{\alpha\sigma'}(x,\vec{r}')
=\bra{\alpha,-k}\hat{h}_T\ket{\sigma',\vec{r}'}$ is a Fourier
transform of the tunneling matrix element.

One can also verify that in the absence of interactions, the
expression for the $\Gamma_{+-}$ amplitude coincides with the
leading term in the zero-bias conductance in the normal state, up to a
replacement $F\mapsto 2(\pi v_F)^{-1}\Im G^R$.  Within a quasiclassical
approximation in the superconductor, \cite{rammer86} one then finds a
relation to the normal-state conductance per unit length, $g(x)$, of the tunnel
interface:
\begin{gather}
  \label{eq:noninteracting-gamma-pm-resistance}
  \Gamma_{+-}(x)
  \simeq
  \frac{1}{4} \hbar v_F R_K g(x)
  =
  \frac{\hbar v_F}{l_T} \frac{R_K}{4 R_N}
  \,.
\end{gather}
Such a relation is typical for NS systems, and connects the amplitude
$\Gamma_{+-}$ to observable quantities. The latter expression assumes
the total resistance $R_N$ is uniformly distributed in a
junction of length $l_T$. When the interface resistance decreases, the
effective pairing amplitude $\Gamma$ grows --- and although not
included in our perturbative calculation, one expects that this
increase is cut off when the effective gap reaches the bulk gap of the
superconductor, $\Gamma_{+-}=\Delta$.

Unlike $\Gamma_{+-}$, the $\Gamma_{++/--}$ amplitudes do not have a
direct relation to the normal-state conductance, and they depend on
the factors $P_{++/--}$, which are proportional to the spin rotation
between the states involved in the pair tunneling (see
Fig.~\ref{fig:pairing-schematics}). Estimating this factor is
necessary for determining how large the same-mode tunneling is in a
given system.

\subsection{Two-particle tunneling}

Making use of the time-reversal symmetry, it is possible to rewrite
the $P$ factors in a more transparent form:
\begin{gather}
  \label{eq:M-single-particle-expression}
  P_{\alpha_1\alpha_2}(k_1,\vec{r}_1';k_2,\vec{r}_2')
  =
  \bra{\alpha_1,-k_1}\hat{Z}(\vec{r}_1',\vec{r}_2')\mathfrak{T}\ket{\alpha_2,-k_2}
  \,,
  \\
  \label{eq:z-operator}
  \hat{Z}(\vec{r}_1',\vec{r}_2')
  \equiv
  \hat{h}_T[1_\sigma\otimes(\ket{\vec{r}_1'}\bra{\vec{r}_2'} + \ket{\vec{r}_2'}\bra{\vec{r}_1'})]\hat{h}_T
  \,,
\end{gather}
where $1_\sigma$ is the identity matrix in the spin space of the
superconductor. Unlike the starting point, this expression is
explicitly independent of the choice of the spin quantization axis.
We also note the symmetry:
\begin{align}
  \label{eq:M-symmetry}
  P_{\alpha_1\alpha_2}(k_1,r_1';k_2,r_2')
  &=
  -
  P_{\alpha_2\alpha_1}(k_2,r_1';k_1,r_2')
  \,,
\end{align}
following from the definition Eq.~\eqref{eq:M-tunneling-element}.

We can now make some remarks on the possibility of $++$ tunneling.
First, suppose that the state $\ket{\alpha,k}$ describes an electron
wave function with a fixed $k$-independent and spatially constant spin
part, and that the tunneling is spin conserving. In this case it is
easy to see that $P_{++}=0$, as the inner product of a spinor and its
time reversed counterpart vanishes. Such a situation is realized, for
instance, within the plain Kane-Mele model. \cite{kane2005-qsh}
Breaking such conditions can, however, lead to $P_{++}\ne0$.  We
demonstrate in the next section that this can occur in HgTe-QW.

\subsection{Effect of Rashba interaction in HgTe/CdTe quantum wells}
\label{sec:hgte-rashba-effect}

We now discuss a simple model for tunneling into the helical edge
states of a HgTe-QW, taking spin axis rotation from the Rashba
interaction into account.  We make the following assumptions: the
tunneling is spin-conserving and local
[$\bra{\vec{r}}\hat{h}_T\ket{\vec{r}'}\propto\delta(\vec{r}-\vec{r}')$]
on the length scales of the four-band model. This results to all
contributions to $P_{++}$ coming solely from the Rashba mixing. While
we cannot estimate the actual values of $P_{+-}$ or $P_{++}$ within
this simplified a model, we can study their relative magnitudes, which
is now determined by the low-energy four-band physics only.

Under the locality and spin-conservation assumptions, the tunnel
matrix element introduced above obtains the following form in terms of
envelope spinor wave functions $\hat{\Psi}$ in the four-band basis
$\{\ket{j}\}=\{\ket{E1+},\ket{H1+},\ket{E1-},\ket{H1-}\}$:
\begin{align}
  \begin{split}
  P_{\alpha_1\alpha_2}(k_1,\vec{r}_1';k_2,\vec{r}_2')
  &=
  \hat{\Psi}_{\alpha_1,-k_1}(x_1',y_1')^\dagger
  \hat{\cal Z}(\vec{r}_1',\vec{r}_2')
  \\
  &\;\times
  \mathfrak{T}
  \hat{\Psi}_{\alpha_2,-k_2}(x_2',y_2')
  + [\vec{r}_1'\leftrightarrow\vec{r}_2']
  \,,
  \end{split}
  \\
  \hat{\cal Z}(\vec{r}_1',\vec{r}_2')_{jj'}
  &=
  \bra{j}h_T[1_\sigma\otimes\ket{\vec{r}_1'}\bra{\vec{r}_2'}]h_T\ket{j'}
  \,.
\end{align}
Time reversal for the four-band spinor reads $\mathfrak{T}=-i\tau_yK$
with $K$ the complex conjugation, and the $\tau$ matrix acts on the
Kramers blocks ($+$, $-$).  For simplicity, we use now a length-scale
separation between the scales appearing in the four-band model
($\Psi$) and the atomic ones (tunneling ${\cal Z}$, $k_{F,S}$ in the
superconductor, unit cell).
We consider only the long-wavelength part
of $P$, and replace ${\cal Z}$ with a constant describing the tunnel
coupling to the quantum well basis states, obtained by averaging it
together with $F$ [cf. Eqs.~\eqref{eq:gamma-plus-plus},
\eqref{eq:gamma-plus-minus-expression}] over $\vec{r}_1'$ and
$\vec{r}_2'$:
\begin{align}
  &\overline{
    [{\cal Z}(\vec{r}_1',\vec{r}_2')+{\cal Z}(\vec{r}_2',\vec{r}_1')]F(\vec{r}_1',\vec{r}_2')
  }
  \notag
  \\
  &\;
  \sim
  \begin{pmatrix}
    {\cal A}(\vec{r}_1')   & {\cal C}(\vec{r}_1')    & 0         & {\cal D}(\vec{r}_1') \\
    {\cal C}(\vec{r}_1')^* & {\cal B}(\vec{r}_1')    & -{\cal D}(\vec{r}_1') & 0 \\
    0          & -{\cal D}(\vec{r}_1')^* & {\cal A}(\vec{r}_1')  & {\cal C}(\vec{r}_1')^* \\
    {\cal D}(\vec{r}_1')^* & 0           & {\cal C}(\vec{r}_1')  & {\cal B}(\vec{r}_1')
  \end{pmatrix}
  F(0)
  \delta(\vec{r}_1'-\vec{r}_2')
  \,,
\end{align}
with ${\cal A}$ and ${\cal B}$ real-valued. This form follows from the
time reversal symmetry and hermiticity of the matrix elements of the
operator in Eq.~\eqref{eq:z-operator}. We have also assumed here that
the decay length for the $F$ function ($\sim{}k_{F,S}^{-1}$) is short
on the scales of the 4-band model.  Finally, we for simplicity neglect
the coupling to the $H1$ band, and set ${\cal B}={\cal C}=0$.  For a
lateral contact (SC on top of HgTe-QW), the main tunnel coupling is
expected to involve the E1 band, which extends deeper
\cite{bernevig2006-qsh} into the CdTe barrier than H1. Including
additional couplings would however cause no essential qualitative
differences in the estimated \emph{ratio} between the $++$ and $+-$
terms.

Without additional spin axis rotation from the Rashba interaction, the
edge states are in separate Kramers blocks (see
Appendix~\ref{sec:edge-states}),
$\hat{\Psi}_{+}\propto{}(\hat{\Phi}_+,0)$ and
$\hat{\Psi}_-\propto{}(0,\hat{\Phi}_-)$, and we can see that
$P_{++}=0$ whereas $P_{+-}\propto{}{\cal A}$.  Note that a
contribution proportional ${\cal D}$ does not arise: the unperturbed
edge state wave functions are both proportional to the same constant
real-valued spinor, $\hat{\Phi}_\pm\propto\hat{\chi}$, so that the
${\cal D}$-dependent contribution would be proportional to
$\hat{\chi}^\dagger i\sigma_y\hat{\chi}=0$.  This structure also
implies that contributions proportional to ${\cal D}$ do not arise in
the leading order of the Rashba coupling.

\begin{figure}
  \includegraphics{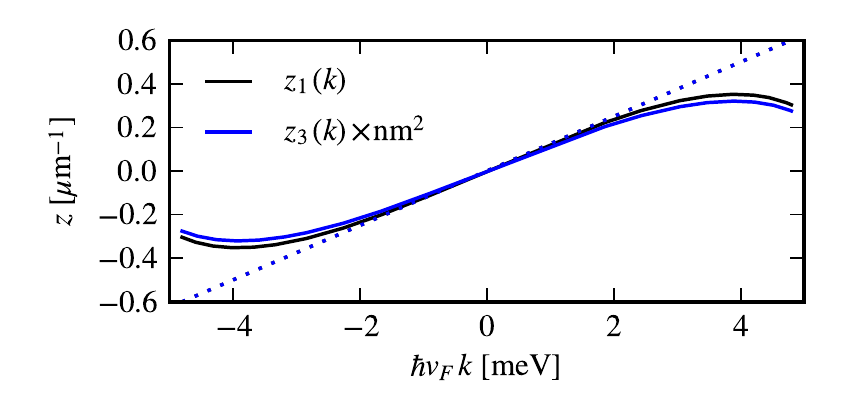}
  \caption{\label{fig:zk-vs-k}
    Projections $z(k)$ of the additional spin-orbit couplings
    $R_0$ and $T_0$ on the edge state basis,
    for the parameters of Ref.~\onlinecite{rothe2010-fod}
    with $M=\unit[-10]{meV}$.
    The dashed lines indicate linear approximations $z_1\approx{}z_{0,1}k$ and
    $z_3\approx{}z_{0,3}k$ with $z_{0,1}=0.03$ and $z_{0,3}=0.03\unit{nm^{-2}}$.
  }
\end{figure}

Rashba and other related spin-orbit interactions in the four-band
model can be represented as \cite{rothe2010-fod}
\begin{align}
  \label{eq:rashba-hamiltonian}
  H_R =
  \begin{pmatrix}
    0 & h_R \\
    h_R^\dagger & 0
  \end{pmatrix}
  \,,
  \quad
  h_R =
  i
  \begin{pmatrix}
    -R_0 k_- & \delta + iS_0 k_-^2 \\
    -\delta - iS_0 k_-^2 & T_0 k_-^3
  \end{pmatrix}
  \,,
\end{align}
where $k_{\pm} = k \pm i k_y$.  For the QW parameters used in
Ref.~\onlinecite{rothe2010-fod}, $R_0\approx\unit[-15.6]{nm^2}\times
eE_z$, $iS_0\approx\unit[-2.10]{nm^3}\times
eE_z$, and $T_0\approx{}\unit[-8.91]{nm^4}\times eE_z$, where $E_z$ is
the electric field perpendicular to the QW plane.  The model could also
include the bulk inversion asymmetry terms
$\delta$. \cite{konig2008-qsh}

To obtain the effect of the Rashba interaction on the wave functions,
we find the low-energy eigenstates of $H=H_0+H_R$ numerically. For
given $k$, this is a 1-D eigenvalue problem in the $y$-direction,
which can be discretized and solved by standard approaches. Analytical
results can be obtained by perturbation theory in $H_R$ restricted to
the low-energy subspace spanned by the unperturbed edge states. For
typical experimental parameters, the whole wave functions $\hat{\Psi}$
however turn out to have a significant component also in the continuum
of bulk modes above the gap, which is not adequately captured by such
an approach. Our estimates for the matrix elements $P_{\alpha\beta}$
below are therefore based on the numerical solutions for the
eigenstates.

However, qualitative understanding can be obtained on the basis of
the model restricted to the low-energy subspace. Projecting $H_R$ to
this basis (see Appendix~\ref{sec:edge-states}), we find the effective
low-energy Hamiltonian of the system \footnote{ Because the analytical
  edge state wave functions have a discontinuous derivative due to the
  boundary condition, the matrix element
  $\int\dd{y}\phi(y)^*\partial_y^3\psi(y)$ is better rewritten as
  $\frac{1}{2}\int\dd{y}[\partial_y^2\phi(y)^*\partial_y\psi(y)-\partial_y\phi(y)^*\partial_y^2\psi(y)]$,
  to remove the need to evaluate boundary terms.}
\begin{align}
  H_R'
  &=
  \begin{pmatrix}
    0 & -i [R_0 w_1(k) + T_0 w_3(k)]
    \\
    \mathrm{c.c.} & 0
  \end{pmatrix}
  \,,
  \\
  w_1(k)
  &=
  \chi_1^2
  \int_{-\infty}^\infty\dd{y}f_{+,k}(y)[k+\partial_y]f_{-,k}(y)
  \\
  w_3(k)
  &=
  \chi_2^2
  \int_{-\infty}^\infty\dd{y}f_{+,k}(y)[k+\partial_y]^3f_{-,k}(y)
  \,,
\end{align}
where $\hat{\Phi}_{\pm,k}(y)=f_{\pm,k}(y)\hat{\chi}$.  This result is
valid to the leading order in $h_R$. The constant and quadratic in $k$
terms (proportional to $\delta$ and $S_0$) give no contribution, as
$\hat{\chi}^\dagger \sigma_y \hat{\chi}=0$. Using typical HgTe-QW
parameters, \cite{rothe2010-fod} the integrals evaluate to
$w_1(k)\approx{}w_{0,1} k$ and $w_3(k)\approx{}w_{0,3}k$ near the
Dirac point, as illustrated in Fig.~\ref{fig:zk-vs-k}.  The prefactor
$w_{0,1}\approx{}0.03$ is essentially independent of the mass
parameter $M$, and
$w_{0,3}\approx{}0.03\unit{nm^{-2}}\times(|M|/\unit[10]{meV})$.  Note
here that the matrix element $0.03R_0k$ of the Rashba interaction with
the edge states is significantly smaller than the $R_0k_\pm$ appearing
in the bulk Hamiltonian.
The $2\times2$ effective Hamiltonian yields the wave functions:
\begin{align}
  \label{eq:rashba-edge-states-analytical}
  \hat{\Psi}_{+,k}
  \simeq
  \begin{pmatrix}
    \hat{\Phi}_{+,k}
    \\
    \frac{iw_0}{2v_F}\hat{\Phi}_{-,k}
  \end{pmatrix}
  \,,
  \quad
  \hat{\Psi}_{-,k}
  \simeq
  \begin{pmatrix}
    \frac{iw_0}{2v_F}\hat{\Phi}_{+,k}
    \\
    \hat{\Phi}_{-,k}
  \end{pmatrix}
  \,,
\end{align}
where $w_0=R_0w_{0,1}+T_0w_{0,3}$.
The Rashba interaction mixes the two Kramers blocks, but in the leading
order does not modify the energy dispersion.
Although the mixing angle of the $\hat{\Phi}_{\pm,k}$
spinors is independent of $k$, the total four-band spinor is not: the
decay lengths $1/\lambda_{1/2}(k,\alpha)$ of $\hat{\Phi}_{\pm,k}$ in
the $y$-direction depend on $k$ and are different for the $\alpha=+$
and $\alpha=-$ states: time-reversal symmetry only guarantees
$\lambda_{1/2}(k,+)=\lambda_{1/2}(-k,-)$.  This makes the electron
spin axis to rotate both spatially and with energy $\varepsilon$,
which ultimately is required for a finite $P_{++}$.

\begin{figure}
  \includegraphics[width=\columnwidth]{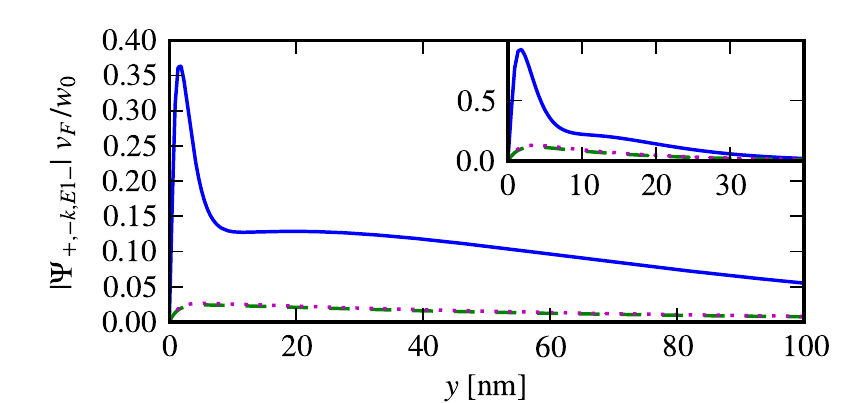}
  \caption{\label{fig:bulk-mode-contribution}
    The $E1-$ component of $|\Psi_{+,-k}|$ at $k=|M|/(6\hbar v_F)$.  
    It is linear in the
    Rashba parameter $w_0/v_F$, provided $|w_0/v_F|\lesssim{}1$.  
    Shown are numerical results (solid
    line), the component in the unperturbed edge state subspace 
    (dashed line), and the result of 
    Eq.~\eqref{eq:rashba-edge-states-analytical} (dotted).  
    The behavior for $y\gtrsim{}\unit[20]{nm}$ depends 
    mainly on the linear in $k$ Rashba term $R_0$.
    Here, $M=\unit[-10]{meV}$.
    Inset: results for $M=\unit[-50]{meV}$.
  }
\end{figure}

For comparison, we show in Fig.~\ref{fig:bulk-mode-contribution} the
$E1-$ component of the numerically computed total edge state wave
function $\Psi_{+,k}(y)$, and its projection to the low-energy
subspace, which can be seen to match
Eqs.~\eqref{eq:rashba-edge-states-analytical} to a very good
accuracy. The $E1-$ component is proportional to the Rashba coupling
and contributes to $P_{++}$. As is clearly visible in the figure,
neglecting the bulk states underestimates the total amount of spin
rotation, for experimentally relevant parameters. For a larger (but
unphysical) value for the gap $|M|$, the low-energy theory works
slightly better, as visible in the inset of
Fig.~\ref{fig:bulk-mode-contribution}.

\begin{figure}
  \includegraphics[width=\columnwidth]{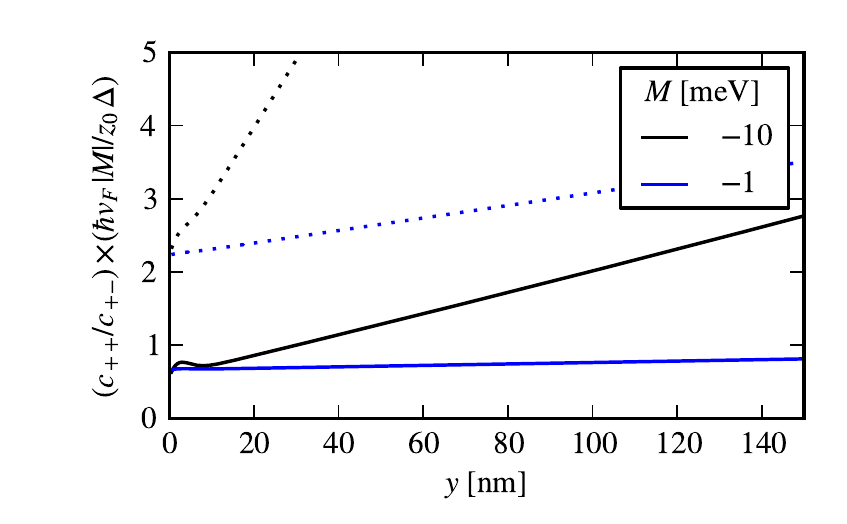}
  \caption{
    \label{fig:gpp-vs-gpm}
    Relative magnitude of the two types of tunneling as a function of
    location and parameters, at $K=-2k_F$ for $P_{++}$ and $K=0$ for
    $P_{+-}$, as obtained from numerically computed $\hat{\Psi}_{\pm,k}$. 
    The scaling with the electric field is given by
    $z_0\equiv\unit[1]{eV\,nm}\times{}E_z/(\unitfrac[500]{mV}{nm})$.
    Solid and dashed lines indicate
    $\hbar{}v_Fk_F=\pm{}|M|/(6\hbar v_F)$, i.e., 
    energies $E = E_{\rm Dirac} \mp |M|/6\approx{}(0.75\mp0.15)|M|$ 
    for the states involved in the $c_{++/+-}$ factors.
    For $z_0/v_F\lesssim{}1$, $c_{++}$ is linear in $z_0$.  
  }
\end{figure}

We can now estimate the relative order of magnitude between $P_{++}$
and $P_{+-}$ within this model.  From the results above, one can see
that the representative quantities to be compared are
\begin{align}
  c_{++}(K,\vec{r}_1',\vec{r}_2')
  &=
  \frac{\Delta}{i \hbar v_F}\partial_k P_{++}(\frac{K}{2}+k,\vec{r}_1';\frac{K}{2}-k,\vec{r}_2')\rvert_{k=0}
  \,,
\end{align}
and
\begin{align}
  c_{+-}(K,\vec{r}_1',\vec{r}_2')
  &=
  P_{+-}(\frac{K}{2}-k_F,\vec{r}_1';\frac{K}{2}+k_F,\vec{r}_2')
  \,.
\end{align}
In Fig.~\ref{fig:gpp-vs-gpm} we show the ratio of these amplitudes for
$\vec{r}_1'=\vec{r}_2'=(0,y)$ (i.e., the value at a distance $y$ from
the edge).  The $c_{++}$ amplitude increases when the energies of the
edge states involved approach the TI energy gap edge. The general
order of magnitude of the $c_{++/+-}$ factors can be estimated to be
of the order
\begin{align}
  \label{eq:order-of-magnitude-estimate}
  c_{++} \sim \frac{\Delta}{|M|} \frac{z_0}{\hbar v_F} c_{+-}
  \,.
\end{align}
A similar relation is then expected also between the $\Gamma_{++}$ and
$\Gamma_{+-}$ factors for surface contacts to area near $y=0$.  Here
and below, we characterize the strength of the Rasbha interaction with
the quantity
$z_0\equiv\unit[1]{eV\,nm}\times{}E_z/(\unitfrac[500]{mV}{nm})$.

Using the above results for the order of magnitude of $c_{++}$ and
$c_{+-}$ we find (see Appendix B) the estimates for the general
case with $e$-$e$ interactions:
\begin{subequations}
\label{eq:gamma-estimates}
\begin{align}
  \Gamma_{+-} &\simeq (a_0\Delta)^{\frac{g+g^{-1}}{2}-1} \Gamma_{+-}^{g=1}
  \\
  \Gamma_{++} &\simeq \frac{1}{2}(a_0\Delta)^{\frac{g+g^{-1}}{2}-1} 
  \frac{z_0\Delta}{v_F |M|}
  \Gamma_{+-}^{g=1}
  \,,
\end{align}
\end{subequations}
corresponding to cutoff $a=\hbar v_F/\Delta$. The
relation~\eqref{eq:noninteracting-gamma-pm-resistance} for the
noninteracting value $\Gamma_{+-}^{g=1}\simeq(\hbar v_F/l_T)R_K/(4R_N)$
fixes the magnitudes relative to experimental parameters.

With finite electron-electron interactions ($g\ne1$) in the helical
liquid, all the effective tunnel rates obtain identical scaling
in the original short-distance cutoff $a_0$. This reflects the
renormalization of the single-particle tunneling elements
$t_{\alpha\sigma}$ by the electron-electron interactions.

We can also estimate the Rashba coupling factor appearing in
$\Gamma_{++}$.  With a typical TI gap
$|M|\sim\unit[10]{meV}\sim\unit[100]{K}$, we see that the factor of
$\Delta/|M|$ can be made of the order of $0.1\ldots0.2$ with
conventional superconductors, and can be even larger for smaller TI
gaps.  The second factor is $z_0/\hbar v_F \sim
E_z/(\unitfrac[100]{mV}{nm})$, and as visible in
Eq.~\eqref{eq:rashba-edge-states-analytical}, measures the rotation of
the spin axis caused by the Rashba mixing.  An upper limit for the
field that can be applied in practice is likely of the order
$E_z\sim\unitfrac[100]{mV}{nm}$, as for fields larger than that, the
potential difference across the QW becomes comparable to the energy
gap of the barrier material (CdTe). Based on this we find an estimate
for the achievable ratio, $\Gamma_{++}\sim{}0.1\ldots 1
\Gamma_{+-}$.

Finally, let us remark that tunneling that is local in real space,
$\bra{\vec{r}}\hat{h}_T\ket{\vec{r}'}\propto{}\delta(\vec{r}-\vec{r}')$,
does not lead to tunneling that is local in the edge state
Hamiltonian, $t_{\alpha\sigma}(x,\vec{r}')\propto\delta(x-x')$.  This
follows in a straightforward way from the extended 2-D nature of the
edge states and the $k_x, k_y$ mixing due to the spin-orbit
interactions:
$\bra{\alpha,x}\hat{h}_T\ket{\sigma,\vec{r}}\propto{}\sum_{k}e^{ik(x-x')}{\cal Y}_{\alpha,k}(\sigma,y',z')^*$.
If the spatial profile ${\cal Y}$ of the wave function has $k$-dependence on the
scale $k_0$, the sum resembles a rounded $\delta$ function of width
$k_0^{-1}$. For HgTe QW edge states, $k_0^{-1}\sim{}\hbar v_F/|M|$ is
a low-energy length scale. Because of this, a pointlike contact to a
superconductor can produce a finite $P_{++}$, even though assuming
$t(x,\vec{r}')\propto\delta(x-x')$  in Eq.~\eqref{eq:M-tunneling-element}
leads to the opposite conclusion.

\section{Transport signatures}
\label{sec:transport}

To study the experimental signatures implied by the above model, we
consider the transport problem in the setups depicted in
Fig.~\ref{fig:setup}.  There, two superconducting contacts are
coupled to a helical liquid, whose potential is tuned by additional
terminals at the ends.  There are three related transport effects one
can study here: the equilibrium dc Josephson effect, the ac Josephson
effect, and the NS conductance.

We consider a general nonequilibrium case of a time-dependent pair
potential $\Delta(t)=|\Delta|e^{i\varphi_1(t)}$ in the left contact
and $\Delta(t)=|\Delta|e^{i\varphi_2(t)}$ in the right one, with
$\varphi_1(t)=\varphi_0/2 + 2V_1t$ and
$\varphi_2(t)=-\varphi_0/2+2V_2t$.  In
Eq.~\eqref{eq:effective-hamiltonian}, the factors $\Gamma$ inherit
this time dependence. We also assume that only sub-gap energies are
involved in the transport, so that the quasiparticle current to the
superconductors remains exponentially suppressed by the
superconducting gap.

The current is obtained as an expectation value of a current operator
$\hat{I}=i[H,\hat{N}]$ where $\hat{N}$ is the particle number in the
HLL. From the effective Hamiltonian, we identify
\begin{align}
  \hat{I} &= \hat{I}_{S1} + \hat{I}_{S2}
  \\
  \hat{I}_{S1}
  &=
  \sum_{\alpha\beta}
  \int_{S1}\dd{x} 2i\Gamma_{\alpha\beta}(x)\psi_\alpha(x)\psi_\beta(x)
  +
  \mathrm{h.c.}
  \,,
  \\
  \hat{I}_{S2}
  &=
  \sum_{\alpha\beta}
  \int_{S2}\dd{x} 2i\Gamma_{\alpha\beta}(x)\psi_\alpha(x)\psi_\beta(x)
  +
  \mathrm{h.c.}
  \,,
\end{align}
where $\hat{I}_{S1}$ and $\hat{I}_{S2}$ must be interpreted as the
parts corresponding to currents injected through the interfaces at
$S1$ and $S2$. The sums over $\alpha\beta$ run over $++$, $+-$, and
$--$.

Considering only the Cooperon terms
[cf. Fig.~\ref{fig:low-energy-reduction}(a)], using perturbation
theory up to second order in $\Gamma$ we find
\begin{align}
  I_{J,S1}(t)
  &=
  -8
  \Im
  \sum_{\alpha\beta}
  \int_{S1}\dd{x_1}\int_{S2}\dd{x_2}
  X_{\alpha\beta}(x_1,x_2)
  \\\notag&\times
  e^{i\varphi_1(t)}
  \int_0^\infty\dd{t'}
  e^{-i\varphi_2(t-t')}
  \Im[\chi_{\alpha\beta}(x_1-x_2,t')]
  \,,
  \\
  I_{J,S2}(t)
  &=
  I_{J,S1}(t)\rvert_{\varphi_1\leftrightarrow\varphi_2}
  \,,
\end{align}
where
\begin{align}
  X_{\alpha\beta}(x_1,x_2)&
  \equiv{}
  e^{2ik_F(\alpha+\beta)(x_1-x_2)}
  \Gamma_{\alpha\beta}(x_1)
  \Gamma_{\alpha\beta}(x_2)^*
  \\
  \chi_{\alpha\beta}(x; t)
  &=
  \frac{
  \avg{
    e^{i\phi_{\alpha}(x,t)}e^{i\phi_{\beta}(x,t)}
    e^{-i\phi_{\alpha}(0,0)}e^{-i\phi_{\beta}(0,0)}
  }_0
  }{(2\pi a)^{2}}
  \,.
\end{align}
The $\alpha\beta=+-$ component of the current coincides with the
result obtained in Ref.~\onlinecite{fazio1995-jct}.  Note that the
terms included here contain the leading order of the dependence in the
phase difference $\varphi_1-\varphi_2$.

The above correlation functions can be evaluated via standard
bosonization techniques: \cite{giamarchi2004-qpi}
\begin{subequations}
\label{eq:cooperon-correlators}
\begin{align}
  \chi_{\alpha\alpha}(x,t)
  &=
  (2\pi a)^{-2}
  B_{\alpha}(x,t)^{g+g^{-1}+2}
  B_{-\alpha}(x,t)^{g+g^{-1}-2}
  \,,
  \\
  \chi_{+-}(x,t)
  &=
  (2\pi a)^{-2}
  B_{+}(x,t)^{1/g}
  B_{-}(x,t)^{1/g}
  \,,
  \\
  B_\pm(x,t)
  &=
  \frac{-iaz}{\sinh[z(ut-ia\mp x)]}
  \,,
\end{align}
\end{subequations}
where $z=\pi T/u$.

In the noninteracting case ($g=1$), we can evaluate the time integrals
analytically, to order ${\cal O}(a^3)$:
\begin{align}
  I_{J,S1}(t)
  &=
  \int_{S1}\dd{x_1}\int_{S2}\dd{x_2}
  [
  j_{J,S1}^{++}
  +
  j_{J,S1}^{--}
  +
  j_{J,S1}^{+-}
  ]
  \\
  j_{J,S1}^{++}
  &=
  \frac{|X_{++}|}{3\pi v_F}
  V_2[(V_2/\Delta)^2+4\pi^2(T/\Delta)^2]
  \\\notag&
  \times
  \cos\Bigl(
  \varphi_0 + 2V_1t - 2V_2(t-|x_1-x_2|/v_F)
  + \phi_0 \Bigr)
  \,,
  \\
  \label{eq:plus-minus-current}
  j_{J,S1}^{--}
  &=
  0
  \,,
  \\
  j_{J,S1}^{+-}
  &=
  -
  \frac{|X_{+-}|}{\pi v_F}
  \frac{2z}{\sinh(2|x_1-x_2|z)}
  \\\notag&
  \times
  \sin\Bigl(\varphi_0 + 2V_1t - 2V_2(t-|x_1-x_2|/v_F)\Bigr)
  \,,
\end{align}
where
$\phi_0(x_1,x_2)\equiv{}4k_F(x_1-x_2)+\arg[\Gamma_{++}(x_1)\Gamma_{++}(x_2)^*]$
is a dynamical phase shift.

Below, we discuss the implications of these results first at
equilibrium and then at finite biases.

\subsection{Equilibrium}

\begin{figure}
  \includegraphics{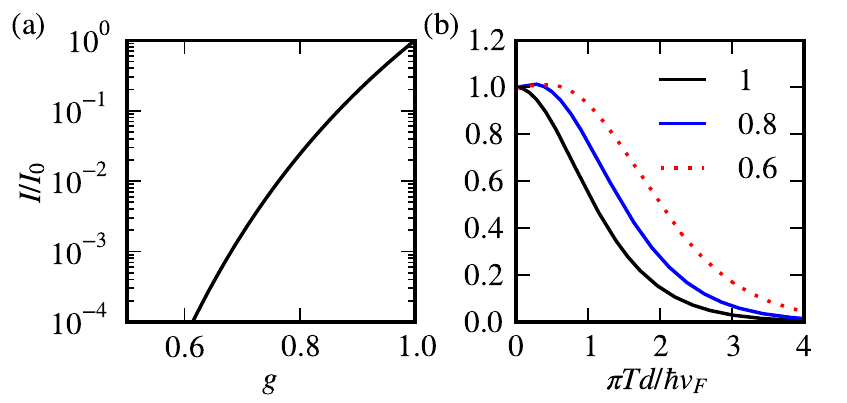}
  \caption{\label{fig:supercurrent}
    Equilibrium Josephson current between the superconducting contacts,
    relative to its noninteracting zero-temperature value $I_0$.
    (a) Dependence $\propto{}\Delta^{2-2/g}$ ($T/\Delta=0.1$) on the
    interaction parameter
    at $T\ll{}\hbar v_F/d$.
    (b) Temperature dependence for different values of $g$.
  }
\end{figure}

At equilibrium, the leading contribution to the supercurrent comes
from the $+-$ channel. As shown in Fig.~\ref{fig:supercurrent},
the supercurrent is finite at zero temperature, and decays
exponentially as the temperature is increased above $\hbar v_F/d$,
in a way that depends on the strength of electron-electron
interactions.  The qualitative features are the same as those found in
Ref.~\onlinecite{fazio1995-jct}.

The contribution from the $++$ and $--$ channels to the equilibrium
current is not more relevant than $+-$ even in the interacting case,
unlike in Ref.~\onlinecite{pugnetti2007-dje}.  Based on scaling
dimensions in the effective Hamiltonian ($\mathrm{dim}\,\psi_+\psi_- =
g^{-1}$, $\mathrm{dim}\,\psi_+\psi_+=g+g^{-1}$), one finds the scaling
$I^{+-}\propto{}(E/\Delta)^{2/g-2}$ and
$I^{++/--}\propto{}(E/\Delta)^{2(g+g^{-1})-2}$ for the low-energy
scale $E=\max(T,v_F/d)$, which implies that $I^{+-}$ will be
more relevant than $I^{++/--}$ whatever the interaction parameter.
This difference arises from the exclusion principle, which makes the
$++/--$ channel less favorable for the supercurrent, although
note that with decreasing $g$ (larger repulsive e-e interaction),
the ($++/--$) contribution grows relative to the $+-$ one. However, as noted
in Section~\ref{sec:low-energy-hamiltonian}, the scaling with the bare
short-distance cutoff $a_0$ as opposed to $\Delta^{-1}$ is identical
for $\Gamma_{++/--}$ and $\Gamma_{+-}$.

\subsection{Nonequilibrium}

When the superconductors are biased with a finite voltage, currents
generically start to flow between all the terminals, and they may also
be time dependent due to the ac Josephson effect. To fully understand
these effects, it is illuminating to compute the spatial distribution
of the currents in the system.

The spatial dependence of the currents in the helical liquid can be
obtained by making use of the following expression for the current
operator $\hat{I}=\frac{v_F}{\sqrt{\pi}}\partial_x\phi(x,t)$ in the
Heisenberg picture (see App.~\ref{sec:current-operator}):
\cite{virtanen2011-dos}
\begin{align}
  \label{eq:heisenberg-current-operator}
  \hat{I}(x,t)
  &=
  \hat{I}_0(x,t)
  +
  v_F
  \int_{-\infty}^\infty\dd{x'}\dd{t'}
  \sum_{\alpha=\pm}\alpha
  \Bigl(
  \\\notag&\quad
    \frac{1+g}{2g}D_{\alpha}(x,t;x',t')
    -
    \frac{1-g}{2g}D_{-\alpha}(x,t;x',t')
  \\\notag&\quad
  \Bigr)
  \hat{j}_{\alpha}(x',t')
  \,,
  \\
  \hat{j}_{\alpha}(x',t')
  &=
  \frac{\delta V(t')}{\delta \phi_\alpha(x')}
  \,.
\end{align}
This applies to any Hamiltonian of the form $H=H_0+V(t)$, where $H_0$
is the bosonized Hamiltonian in Eq.~\eqref{eq:bosonized-hamiltonian};
$\hat{I}_0$ is the current operator evolving in time with the unperturbed
Hamiltonian $H_0$.  The operator $j_\alpha(x',t')$ can be interpreted
as the current density injected to the mode $\alpha=\pm$ at
position $x'$ at time $t'$. The functions $D_{+(-)}$ are initially
right(left)-propagating $\delta$ pulses originating at point $x'$ at
time $t'$.

Let us for simplicity assume that the two superconducting contacts are
pointlike in the low-energy model, that $g=1$, and that the helical
liquid is homogeneous. Then,
$D_\pm(x,t;x',t')=\theta(t-t')\delta(x-x'\mp v_F(t-t'))$ and we find
\begin{align}
  \label{eq:current-result}
  I(x,t) &= \sum_{j=1,2}\bigl[
  \theta(x-x_j)l_T\avg{\hat{I}_{+,j}(t-\frac{|x-x_j|}{v_F})}
  \\\notag&\qquad
  -
  \theta(x_j-x)l_T\avg{\hat{I}_{-,j}(t-\frac{|x-x_j|}{v_F})}
  \bigr]
  \,.
  \\\notag
  \hat{I}_{+,j}
  &=
  \frac{\partial}{\partial\phi_+}\Bigl[
  \Gamma_{++}\psi_+\psi_+
  +
  \Gamma_{--}\psi_-\psi_-
  \\&\qquad
  +
  \Gamma_{+-}\psi_+\psi_-
  +
  \mathrm{h.c.}
  \Bigr]\rvert_{x=x_j}
  \\\notag
  &=
  2i
  \Gamma_{++}\psi_+\psi_+
  +
  i
  \Gamma_{+-}\psi_+\psi_-
  +
  \mathrm{h.c.}
  \,,
  \\
  \hat{I}_{-,j}
  &=
  2i
  \Gamma_{--}\psi_-\psi_-
  +
  i
  \Gamma_{+-}\psi_+\psi_-
  +
  \mathrm{h.c.}
  \,,
\end{align}
where $l_T$ is the small contact size, and the expectation values
$\avg{\cdot}$ closely correspond to the different parts of the
injection currents $I_{J,S1/S2}$ evaluated in the previous section.
Indeed,
\begin{subequations}
\begin{align}
  \avg{\hat{I}_{+,1}(t)} &= j^{++}_{S1}(t) + \frac{1}{2}j^{+-}_{S1}(t) \,, \\
  \avg{\hat{I}_{-,1}(t)} &= j^{--}_{S1}(t) + \frac{1}{2}j^{+-}_{S1}(t) \,, \\
  \avg{\hat{I}_{+,2}(t)} &= j^{++}_{S2}(t) + \frac{1}{2}j^{+-}_{S2}(t) \,, \\
  \avg{\hat{I}_{-,2}(t)} &= j^{--}_{S2}(t) + \frac{1}{2}j^{+-}_{S2}(t) \,.
\end{align}
\end{subequations}
The physical interpretation is particularly simple: the contacts at
$x_1$ and $x_2$ inject current to the helical liquid.  The component
due to $+-$ tunneling splits evenly to the left and right-moving
modes, whereas the $++$ and $--$ components end up solely in the
$+$ and $-$ modes, respectively.  Within each edge mode, the
injected current propagates with the Fermi velocity, as indicated by
the retarded time arguments.

The calculations done in the previous section indicated that in this
case $j_{S1}^{--}=0$ and $j^{++}_{S2}=0$ to leading order in
$a$. Therefore, essentially all of the current injected by the $++$
and $--$ tunneling in fact flows only to the reservoirs that maintain
the chemical potential of the helical liquid at $\mu=0$, rather than
between the two superconducting contacts, which can be verified by
computing the current at $x<x_1,x_2$ and at $x>x_1,x_2$.  The effect
essentially amounts to a modulation of the NS conductance between the
superconductors and the normal leads by the (time-dependent) phase
difference $\varphi_1(t)-\varphi_2(t)$ between the superconducting
contacts.

Based on the above results, we can write down an expression for the
part of the NS current [see Fig.~\ref{fig:setup}(c)] that depends
on the phase difference, in the configuration $V_1=V_2=V$:
\begin{align}
  \label{eq:ns-current}
  \delta I_{NS}
  &=
  \frac{2l_T^2|X_{++}|}{3\pi}
  V\cos(2Vd + \phi_0)\cos(\varphi_0)
  \\\notag
  &\qquad\times
  [(V/\Delta)^2 + 4\pi^2(T/\Delta)^2]
  \\\notag
  &+
  \frac{4l_T^2|X_{+-}|}{\pi}
  \sin(2Vd)\cos(\varphi_0)\frac{z}{\sinh(2zd)}
  \,.
\end{align}
Note that the modulation of the NS conductance from the $+-$ channel
decays exponentially as the temperature increases, whereas the $++$
contribution does not. The same situation should persist in all orders
of perturbation in the effective Hamiltonian for the $+-$ tunneling:
the terms coupling to $\varphi_0$ contain inequal numbers of
$\psi_+(d)\psi_-(d)$ and $\psi_+^\dagger(d)\psi_-^\dagger(d)$, which
implies that the correlation function is of the form
$[B_+(d,t)B_-(d,t)]\times{\cal O}(1)$ and thus has an overall
exponential prefactor $e^{-2\pi Td/v_F}$.  Therefore, there in
principle is a temperature regime at $T\gg{}\hbar v_F/k_B d$ in which
the leading contribution to the $\varphi_0$ dependence of the NS
current comes mainly from the $++$ tunneling, despite the power-law
suppression of this channel in helical liquids.  The physical reason
for the difference can be seen in Fig.~\ref{fig:pairing-schematics}:
for the $+-$ channel an electron pair injected to energies
$\pm\varepsilon$ and traversing through the junction obtains an energy
dependent phase factor $e^{i(k_1+k_2)d}=e^{-2i\varepsilon{}d/v_F}$,
typical of Andreev reflection, which averages towards zero when a
finite energy window is considered.  For the $++$ channel, because of
the linear spectrum, the corresponding phase factor $e^{2ik_Fd}$ is
energy-independent and no such averaging occurs.

The above result requires validity of the perturbation theory, i.e.,
$l_T\Gamma_{+-}\sim{}R_K/4R_N\lesssim{}1$ where $l_T$ is the contact
length.  If this condition is not satisfied, an additional
contribution decaying only as $1/T$ in temperature arises in the $+-$
channel.  \footnote{This can be checked using the Bogoliubov--de
  Gennes equation.}  This additional proximity effect contribution is
similar to what occurs in metallic systems of a more macroscopic size,
\cite{pothier1994-fac,*hekking1993-iot,*volkov1997-lpe} although there
it can be much amplified as the electrons can stay a long time near
the NS interface due to impurity scattering.

With finite repulsive interactions ($g<1$), also the $+-$ contribution
to the NS conductance obtains a power-law prefactor according to the
scaling dimensions, $I^{+-}\propto{}(E/\Delta)^{2/g-2}$ with
$E=\max(T,V)$, and the prefactor of the $++$ part is modified,
$I^{++/--}\propto{}(E/\Delta)^{2(g+g^{-1})-2}$.  According to the
correlation functions \eqref{eq:cooperon-correlators}, exponential
decay will also appear in the $++$ part due to charge
fractionalization, but it will be weaker than in the $+-$ part for all
values of $g$.

A second distinguishing feature of the $++$ contribution to the NS
current is that it is expected to oscillate not only as a function of
the bias, but also as a function of the Fermi wave vector appearing in
the dynamical phase $\phi_0$. In metals or other systems where $k_F$
is large, the wavelength of such oscillations would be on the atomic
length scales, and the contribution would average to zero [as
$\sim\sinc(k_F w)^2$] over any practical contact size
$w$. \cite{pugnetti2007-dje} However, this needs not be the case in
HgTe-QW (or in nanotubes, see Ref.~\onlinecite{pugnetti2007-dje})
when the Fermi level lies close to the Dirac point: for example
assuming $|\mu-E_{\rm Dirac}|\sim{}M/6\sim\unit[1.5]{meV}$ one finds
$1/k_F\sim\unit[150]{nm}$. Such length scales are likely
experimentally accessible.

One should also note that the finite wave velocity combined with the
ac Josephson effect causes some additional effects. The current
propagates at the (renormalized) Fermi velocity $v_F/g$, rather than
at the substantially higher speed of light $c$ at which
electromagnetic excitations propagate.  Assuming only the $+-$ channel
contributes, one can find the spatial dependence of current between
the two contacts:
\begin{align}
  I(t) &\propto
  \sin[2Vt - \frac{2Vx}{v_F}]
  +
  \sin[2Vt - \frac{2V(d-x)}{v_F}]
  \,.
\end{align}
Based on this, it is clear that for biases $V\gtrsim\hbar v_F/ed$
between the two superconducting electrodes, the ac Josephson effect
must be associated with appreciable standing wave oscillations in the
charge density.  This behavior is not specific to helical liquids: a
similar spatially resolved calculation as above for the spinful liquid
ac Josephson effect of Ref.~\onlinecite{fazio1995-jct} should also
produce this feature.  Whether such effects are observable in reality,
however, depends on how realistic the model assumptions about
screening are in the systems studied (see also
Ref.~\onlinecite{egger1998-avs}).

\section{Discussion and conclusions}
\label{sec:discussion}

In this work we considered the proximity effect induced in a helical
edge state, taking into account a spatially and energetically
non-constant spin quantization axis. Such rotation of the spin axis
naturally arises from the spin-orbit interaction in real materials
such as the HgTe QWs, for example in a controlled way by structure
inversion symmetry breaking Rashba terms. This has the consequence
that the singlet correlations in an s-wave superconductor can also
induce a proximity effect in the same channel of left and right-movers
($\Gamma_{++/--}$) in addition to the usual term where the correlation
is between opposite chiral states (with amplitude $\Gamma_{+-}$). We
derived a description of the proximity effect in both channels in the
presence of Rashba interaction using a simple model for the tunneling
between the superconductor and the helical edge state, respecting spin
conservation and time reversal symmetries.

The extra transport channels ($++/--$) describe processes that are in
principle parasitic for the splitting of a Cooper pair into two electrons
propagating into different directions (the $+-$ channel)
\cite{sato2010-cii}. For a single superconducting contact to the
helical liquid, the scaling with temperature (or bias voltage) at low
energies however always favors the $+-$ channel. In
Ref. \onlinecite{recher2002-sct}, the two-particle tunneling into the
bulk of a spinful Luttinger liquid was found to be suppressed in a
power law in $1/\Delta$ similarly as here, but there the tunneling
into the $++/--$ channel was found to be dominant. The difference
arises because in a spinful liquid the two opposite spins can tunnel
into different spin channels, and therefore no Pauli-blocking factors
appear.

Observing effects related to the same-mode tunneling ($\Gamma_{++}$)
is likely rather challenging, as they can be suppressed relative to
$\Gamma_{+-}$ by several factors: the power law suppression
$(T/\Delta)^2+(V/\Delta)^2$ from exclusion principle, suppression of
the tunneling factor $\Gamma_{++}$ itself, and averaging effects
related to contacts if they are larger than $1/k_F$
(i.e. $\sim\unit[150]{nm}$ for parameters in
Fig.~\ref{fig:gpp-vs-gpm}).  However, by observing the dependence of
the NS conductance on the superconducting phase, the relative
difference can be reduced due to the exponential dephasing of the $+-$
contribution at high temperatures. (In the case that the only mode of
transport is via the $+-$ channel, the modulation would still contain
features distinct to ballistic transport, such as oscillations as the
bias voltage is increased.) The question is therefore more on how
small signals can be detected in the conductance, oscillating with the
phase difference $\varphi_0$, and how large the thermal factor $2\pi
k_B T d/\hbar v_F$ can be made before inelastic interaction effects
(e.g. electron-phonon scattering), which we have neglected, start to
play a role.

We find that controlling the spin axis via external electric fields in
HgTe-QW in general requires very strong fields, because of the weak
coupling of the additional spin-orbit interactions to the edge states.
For our case, this makes achieving a large $\Gamma_{++}$ more
difficult, and may in general pose problems to proposals relying on
the control of the spin axis.  Making an optimistic estimate, we find
from Eqs.~\eqref{eq:gamma-estimates} and \eqref{eq:ns-current} that
the ratio of the two contributions to the amplitude of phase-dependent
oscillations in the conductance is ($V\ll{}T$, $g=1$,
$E_z\sim{}\unitfrac[100]{mV}{nm}$)
\begin{align}
  \frac{\delta G_{++}(\varphi_0)}{\delta G_{+-}(\varphi_0)}
  &=
  \frac{2\pi^2}{3}
  \left|\frac{\Gamma_{++}}{\Gamma_{+-}}\right|^2
  \left(\frac{k_BT}{\Delta}\right)^2
  \sinhc\left(\frac{2\pi k_B T d}{\hbar v_F}\right)
  \\\notag
  &\sim
  \left(\frac{k_B T}{M}\right)^2
  \sinhc(2\pi k_B T d/\hbar v_F)
  \,,
\end{align}
where $\sinhc(x)=\sinh(x)/x$. With finite $e$-$e$ interactions
[cf. Eq.~\eqref{eq:gamma-interacting-magnitudes}], the ratio is
multiplied by $(\Delta/T)^{2-2g}$, making the result depend only on $\Delta/M$
in the limit $g\to0$, and the exponential dependence
becomes $\sim\exp(2\pi g[2 - g]k_B T d/\hbar v_F)$. Taking
junction length $d=\unit[3]{\mu m}$, the temperature scale of the
exponential suppression factor is $E_T\equiv\hbar v_F/(2\pi
d)\approx\unit[0.2]{K}$, and the ratio becomes unity at the cross-over
temperature $T_*\approx
E_T\{\log[2(M/E_T)^2]-\log\log[2(M/E_T)^2]\}\approx{}\unit[2.2]{K}$,
which depends weakly on $M$ (here $M=\unit[-10]{meV}$). Given a
suitable superconducting material, this should be achievable.

Another option for amplifying the same-mode tunneling could be to
break the time-reversal symmetry and introduce additional spin flips
or spin rotation, for example via magnetic impurities or ferromagnets.
The effect could still be detected in the NS conductance, as that
conclusion is only based on the generic form of the low-energy
effective Hamiltonian.

Observe that in our analysis the true 2D nature of the edge states in
HgTe-based QWs was important.  The ($++/--$) proximity channel cannot
be found in a completely 1D description, as in such a picture the spin
quantization axis is simply rotated {\it globally} by the Rashba terms
(cf.  Refs.~\onlinecite{sato2010-cii,vayrynen2011-ema}). Such
rotations can have no consequences for Cooper pair injection into a
single edge, due to the $s$-wave symmetry of the pairing
[cf. Eq.~\eqref{eq:M-single-particle-expression}].  Spatially
inhomogeneous Rashba interaction, \cite{strom2010-edi} could, however,
induce a finite $++/--$ amplitude.

As in other systems with small critical currents, \cite{fazio1995-jct}
also here thermal fluctuations in the superconducting phase difference
are a problem for measurements of the temperature-dependence of the
Josephson effect: the temperature scale relevant for the phase
fluctuations, $E_J=\hbar I_c/2e$, is smaller than the intrinsic one,
$E_T=\hbar v_F/d$. More complicated measurement schemes
\cite{golubov2004-cri,dellarocca2007-moc} than the simple
current-biased setup in Fig.~\ref{fig:setup}(b) may nevertheless
help in overcoming this problem.  One should, however, note that only
the Josephson current is a problematic observable in this respect. The
measurement of phase oscillations of the NS conductance in the setup
of Fig.~\ref{fig:setup}(c) is expected to suffer much less from
phase fluctuations, as there the phase difference is locked by the
magnetic flux and the large critical current of the superconducting
loop itself.

In summary, starting from a tunneling Hamiltonian, we derived an
effective low-energy theory describing the superconducting proximity
effect in the helical edge state of a 2D topological insulator.  We
showed that in these systems, despite the s-wave symmetry of the
superconductor, correlations can occur both in ($++/--$) and between
($+-$) the left and right moving modes, and within a simple model, we
estimated the expected magnitudes for the effective proximity gap
parameters in HgTe/CdTe quantum wells.  Based on the effective
Hamiltonian, we studied the dc and ac Josephson effects in the helical
liquid, and considered phase-dependent oscillations of the NS
conductance. In nonequilibrium, we found that correlations within the
same mode can give rise to a long-ranged interference effect, which
could act as a signature of their presence.  Our results also shed
light on the meaning of "spin" in the helicity of these edge states
which is of importance if one intends to use these edge states for
spin-injection or spin-detection.

\acknowledgements

We thank H.~Buhmann, C.~Br\"une, F.~Dolcini, L.~Molenkamp, E.G.~Novik,
and B.~Trauzettel for useful discussions.  We acknowledge financial
support from the Emmy-Noether program of the Deutsche
Forschungsgemeinschaft and from the EU--FP7 project SE2ND.

\appendix

\section{HgTe/CdTe QW edge states}
\label{sec:edge-states}

The edge states of a HgTe-QW can be described within the four-band
model introduced in Ref.~\onlinecite{bernevig2006-qsh}.  Here, we
derive explicit analytical expressions for the edge states in a single
edge following the approach of Ref.~\onlinecite{zhou2008-fse}, for use
in Section.~\ref{sec:hgte-rashba-effect}, and to demonstrate that the
direction where the 4-band spinors point is independent of $k$ and
$M$, in the absence of inversion symmetry breaking terms.

The four-band Hamiltonian reads
\begin{align}
  H &= \begin{pmatrix} h(k) & 0 \\ 0 & h(-k)^* \end{pmatrix}
  \,,
  \\
  h(k) &= \epsilon(k)\sigma_0 + \vec{d}(k)\cdot\vec{\sigma} \,,
  \;
  \epsilon(k) = C - D |\vec{k}|^2 \,,
  \\
  \vec{d}(k) &= (A k, -A k_y, M - B k^2)
  \,.
\end{align}
For the parameters $A$, $B$, $C$, $D$ we use values from
Ref.~\onlinecite{rothe2010-fod}: $A=\unit[365]{meV\,nm}$,
$B=\unit[-706]{meV\,nm^2}$, $D=\unit[-532]{meV\,nm^2}$ and take $C=0$
(it only shifts the Dirac point). For an edge with QW lying at $y>0$,
with the wave function vanishing at $y=0$, the edge eigenstates are:
\begin{align}
  \hat{\Psi}_{+,k} &= \begin{pmatrix}\hat{\Phi}_{+,k} \\ 0 \end{pmatrix}
  \,,
  \;
  \hat{\Psi}_{-,k} = \begin{pmatrix}0\\\hat{\Phi}_{-,k}\end{pmatrix}
  \,,
  \\
  \hat{\Phi}_{\alpha,k}
  &=
  N
  e^{-ik x}
  \bigl[
    e^{-\lambda_1 y}
    -
    e^{-\lambda_2 y}
  \bigr]
  \begin{pmatrix}
    -\sqrt{D - B}
    \\
    \sqrt{-D - B}
  \end{pmatrix}
  \,,
\end{align}
where
\begin{align}
  E
  &=
  \frac{-DM}{B} - \alpha k \frac{A\sqrt{B^2-D^2}}{B}
  \,,
  \\
  \lambda_{1/2} &= \sqrt{k^2 + F \mp \sqrt{F^2 - Q^2}}
  \,,
  \\
  F &= \frac{A^2 - 2(BM+DE)}{2(B^2 - D^2)}
  \,,
  \;
  Q^2 = \frac{M^2 - E^2}{B^2 - D^2}
  \,,
\end{align}
and $N$ is a normalization constant.

Note that the spinor points to a single direction independent of $k$
or energy $E$, so that the above results are of the form
\begin{align}
  \hat{\Psi}_{+,k} =
  \begin{pmatrix}\hat{\chi} \\ 0\end{pmatrix}
  f_{+,k}
  \,,
  \;
  \hat{\Psi}_{-,k} =
  \begin{pmatrix}0 \\ \hat{\chi}\end{pmatrix}
  f_{-,k}
  \,,
\end{align}
where the constant spinor $\hat{\chi}$ is normalized
($\hat{\chi}^\dagger\hat{\chi}=1$) and depends only on the parameters
$B$ and $D$. The envelope $f_{\alpha,k}(x,y)=N'e^{ik
  x}[e^{-\lambda_1 y}-e^{-\lambda_2 y}]$ is a scalar function.  Using
the above parameters we find $\hat{\chi}=(0.35, 0.94)^T$, so that
the spinor has the main contribution in the $H1$ band.

\section{Low-energy Hamiltonian}

\label{sec:low-energy-hamiltonian}

\begin{figure}
  \includegraphics{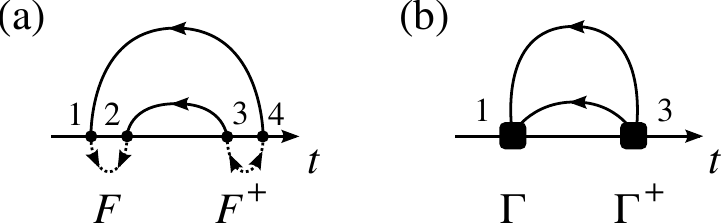}
  \caption{
    \label{fig:low-energy-reduction}
    (a) Cooperon giving a contribution to the Josephson current.
    (b) Integrating out the relative coordinates gives
    an effective description of Andreev reflection.
  }
\end{figure}

In this Appendix, we derive the effective Hamiltonian in
Eq.~\eqref{eq:effective-hamiltonian} via perturbative renormalization
group (RG), \cite{giamarchi2004-qpi} aiming to approximate the
Cooperon term [see Fig.~\ref{fig:low-energy-reduction}(a)] appearing
in the perturbation expansion in the tunneling with the effective term
in Fig.~\ref{fig:low-energy-reduction}(b).  When computing the
Josephson current, for the $\Gamma_{+-}$ channel this approach is
compatible with that used in Ref.~\onlinecite{fazio1995-jct} and
elsewhere in the long-junction case. For the $\Gamma_{++}$ and
$\Gamma_{--}$ channels we however need to pay more attention to the
tunneling elements.

We take our effective Hamiltonian to have the form:
\begin{align}
  H_{\rm eff}
  &=
  H_0 + H_T + H_{T2}
  =
  H_0 + V
  \,,
  \\
  H_T &=
  \sum_{\alpha=\pm, \sigma'=\uparrow,\downarrow}
  \int\dd{x}\dd{^3r'} t_{\alpha\sigma'}(x,\vec{r}')
  \psi^\dagger_{\alpha}(x)\psi_{S\sigma'}(\vec{r}')
  +
  \mathrm{h.c.}
  \\
  H_{T2}
  &=
  \sum_{\alpha\beta}
  \int\dd{x} \Gamma_{\alpha\beta}(x) \psi_\alpha(x) \psi_\beta(x+a)
  +
  \mathrm{h.c.}
  \,,
\end{align}
and rescale only the cutoff in the helical liquid in the progress of
RG. Because correlations in the superconductor decay exponentially at
distances $\gtrsim\Delta^{-1}$, and because the superconducting gap
prohibits dissipation at low energies inside the superconductor, $H_T$
can be neglected in calculations after the scaling to low-energy
length scales $\gg{}\Delta^{-1}$ is done --- which reduces the
effective Hamiltonian to that in Eq.~\eqref{eq:effective-hamiltonian}.

The scaling equations read
\begin{align}
  \frac{\dd{t_{\alpha\sigma'}}}{\dd{l}}
  &= [2 - \eta_1] t_{\alpha\sigma'}
  \\
  \frac{\dd{\Gamma_{\alpha\beta}}}{\dd{l}}
  &= [2 - \eta_{2,\alpha\beta}] \Gamma_{\alpha\beta} + S_{\alpha\beta}(l)
  \,,
\end{align}
where $\eta_1 = (g+g^{-1})/4$ is the scaling dimension of
$e^{i\phi_\pm}$ appearing in tunneling $H_T$, and
$\eta_{2,\alpha,-\alpha} = 1/g$ and $\eta_{2,\alpha,\alpha} = g +
g^{-1}$ are scaling dimensions of the operators in $H_{T2}$.  Although
essentially a standard calculation, below we explain the derivation of
$S_{\alpha\beta}$ in detail.

Below, we need the factorization
\begin{gather}
  \label{eq:boson-fermi-expansion}
  \psi_{\alpha_1}(x_1,\tau_1)
  \psi_{\alpha_2}(x_2,\tau_2)
  =
  (a_0 q_0)^{2\eta_1}
  \frac{U_{\alpha_1}(\tau_1)U_{\alpha_2}(\tau_2)}{2\pi a_0}
  \\\notag
  \times
  e^{ik_F(\alpha_1 x_1 + \alpha_2 x_2)}
  \\\notag
  \times
  {:e^{i\phi_{\alpha_1}(x_1,\tau_1)}e^{i\phi_{\alpha_2}(x_2,\tau_2)}:}
  \,
  C_{\alpha_1,\alpha_2}(x_1-x_2, \tau_1-\tau_2)
  \,,
\end{gather}
where $q_0=2\pi/L$ is the infrared cutoff, and the correlation
functions read
\begin{align}
  C_{++}(z)
  &=
  (q_0z)^{g^{-1}(\frac{1+g}{2})^2}
  (q_0z^*)^{g^{-1}(\frac{1-g}{2})^2}
  \\
  C_{\alpha,-\alpha}(z)
  &=
  |q_0 z|^{(1-g^2)/(2g)}
  \,,
\end{align}
where $C_{\alpha\beta}(x,\tau)=C_{\alpha\beta}(z)$, $z=v_F\tau - ix$, and
$C_{--}(z) = C_{++}(z)^*$.  Observe that
$C_{\alpha_1\alpha_2}(z)\propto
q_0^{\eta_{2,\alpha_1\alpha_2}-2\eta_1}$.

To perform the RG steps, we also need the corresponding operator product
expansions. Taking sign changes due to Klein factors and time ordering
into account, we find [cf. Eq.~\eqref{eq:boson-fermi-expansion}]:
\begin{gather}
  \label{eq:ope}
  T[
  \psi_{\alpha_1}(z_1)
  \psi_{\alpha_2}(z_2)
  ]
  \\\notag
  \simeq
  u_{\alpha_1\alpha_2}(z_1-z_2)
  \psi_{\alpha_1}(\frac{z_1+z_2}{2})\psi_{\alpha_2}(\frac{z_1+z_2}{2} + a_0)
  \,,
\end{gather}
where
\begin{align}
  u_{\alpha_1\alpha_2}(z)
  =
  e^{i(\alpha_1-\alpha_2) k_F x/2}
  \sgn(\tau)^{\delta_{\alpha_1,\alpha_2}}
  \\\notag
  \times
  \frac{C_{\alpha_1,\alpha_2}(x\sgn(\tau),|\tau|)}{C_{\alpha_1,\alpha_2}(0,a_0)}
  \,,
\end{align}
and $\delta_{\alpha_1,\alpha_2}$ in the sign factor arises from
the fact that
$U_{\alpha_1}U_{\alpha_2}=(-1)^{1+\delta_{\alpha_1,\alpha_2}}U_{\alpha_2}U_{\alpha_1}$.

The source term $S_{\alpha\beta}(l)$ for the Andreev reflection
processes $\Gamma_{\alpha\beta}$ appears from the second-order term in
the pertubation expansion of the partition function, $Z/Z_0=\avg{T
  e^{-\int_0^\beta \dd{\tau}\lambda
    V(\tau)}}_0=1+c_1\lambda+c_2\lambda^2+\ldots$.
Combining two $H_T$ and using the operator
product expansions gives a contribution to $\Gamma_{\alpha\beta}$.  We
also trace out the superconductors at this step, factorizing the
expectation value to
$\avg{\ldots}_0=\avg{\ldots}_{HLL,0}\avg{\ldots}_{S,0}$.  This yields
the result
\begin{align}
  \frac{\dd{c_2}}{\dd{l}}
  &=
  \int\dd{^2z}
  \sum_{\alpha\beta}
  \avg{T[
    \psi_{\alpha}(z)\psi_{\beta}(z + a_0)
  ]}_0
  S_{\alpha\beta}(l,z)
  \,,
  \\
  S_{\alpha\beta}(l,z)
  &=
  a_0 [\partial_r f_{\alpha\beta}(l,z,r)]_{r=a_0}
  \,,
  \\
  \label{eq:f-expression}
  f_{\alpha\beta}(l,z,r)
  &=
  \int_{|z'|<r}\dd{^2z'}
  c_{\alpha\beta}(l,z,z')
  \,,
  \\
  \label{eq:c-expression}
  c_{\alpha\beta}(l,z,z')
  &=
  \frac{1}{2}
  \int\dd{^3r_1'}\dd{^3r_2'}
  \sum_{\sigma_1\sigma_2=\uparrow,\downarrow}
  \\\notag
  &
  \times
  t_{\alpha\sigma_1}(x+x'/2,\vec{r}_1')^*
  t_{\beta\sigma_2}(x-x'/2,\vec{r}_2')^*
  \\\notag
  &
  \times
  F^\dagger(\sigma_1,\vec{r}_1',\tau';\sigma_2,\vec{r}_2',0)
  u_{\alpha\beta}(l,z')
  \,,
\end{align}
where $\dd{^2z}$ is shorthand for $\dd{x}\dd{\tau}$.
This closes the set of equations.

We can now solve the scaling equations:
\begin{align}
  t_{\alpha\sigma'}(l) &= e^{(2-\eta_1)l}t_{\alpha\sigma}(0)
  \,,
  \\
  \Gamma_{\alpha\beta}(l,z)
  &=
  \int_0^{l}\dd{s}
  e^{(2-\eta_{2,\alpha\beta})(l-s)}
  S_{\alpha\beta}(s,e^{l-s}z)
  \,.
\end{align}
The integral appearing in $\Gamma_{\alpha\beta}$ can be simplified by
substituting in the scaling obtained for $t_{\alpha\beta}$, and
undoing the rescaling of length scales in the remaining integrals.
This yields:
\begin{align}
  u_{\alpha\beta}(l,z)
  &=
  e^{[2\eta_1 - \eta_{2,\alpha\beta}]l}
  u_{\alpha\beta}(0, e^l z)
  \\
  c_{\alpha\beta}(l,z,z')
  &=
  e^{[4-\eta_{2,\alpha\beta}]l}
  c_{\alpha\beta}(0,e^{l}z,e^{l}z')
  \\
  f_{\alpha\beta}(l,z,r)
  &=
  e^{[2-\eta_{2,\alpha\beta}]l}f_{\alpha\beta}(0,e^{l}z,e^{l}r)
  \,,
  \\
  S_{\alpha\beta}(l,z)
  &=
  e^{[3 - \eta_{2,\alpha,\beta}]l}
  a_0
  [\partial_r
  f_{\alpha\beta}(0,e^l z, r)]_{r=a_0 e^l}
\end{align}
And further,
\begin{align}
  \Gamma_{\alpha\beta}(l,z)
  &=
  e^{[2-\eta_{2,\alpha\beta}]l}
  \int_{a_0<|z'|<a(l)}\dd{^2z'}
  c_{\alpha\beta}(0, e^l z, z')
  \,,
\end{align}
where $a(l)=a_0 e^l$. Here, $c_{\alpha\beta}$ decays
fast for $|z'|>|\Delta|^{-1}$ due to the decaying $F$ functions and
the assumedly short range of tunneling. Therefore, at long length
scales $a=a_0 e^l \gg{}|\Delta|^{-1}$ we can replace the upper limit
in the integral: $a\mapsto{}\infty$.

Undoing all length rescaling, we can write the result in the form of
Eq.~\eqref{eq:effective-hamiltonian}, with
\begin{align}
  \label{eq:effective-gamma-general}
  \Gamma_{\alpha\beta}(x)
  &=
  \frac{1}{4}
  \int\dd{^3r_1'}\dd{^3r_2'}\dd{x}'\dd{\tau}'
  e^{i(\alpha-\beta) k_F x'/2}
  \\\notag
  &
  \times
  P_{\alpha\beta}(x+\frac{x'}{2}, \vec{r}_1'; x-\frac{x'}{2}, \vec{r}_2')^*
  F^\dagger(\vec{r}_1',\tau';\vec{r}_2',0)
  \\\notag
  &
  \times
  \frac{
    a^{1-\eta_{2,\alpha\beta}}
    C_{\alpha\beta}(x'\sgn(\tau'),|\tau'|)
    \sgn(\tau')^{\delta_{\alpha\beta}}
  }{
    a_0^{1-\eta_{2,\alpha\beta}}
    C_{\alpha\beta}(0,a_0)
  }
  \,,
\end{align}
where
\begin{align}
  \label{eq:M-tunneling-element-x-space}
  P_{\alpha_1\alpha_2}(x_1,\vec{r}_1';x_2,\vec{r}_2')
  &\equiv
  [
  t_{\alpha_1\downarrow}(x_1,\vec{r}_1')t_{\alpha_2\uparrow}(x_2,\vec{r}_2')
  \\
  \notag
  &
  -
  t_{\alpha_1\uparrow}(x_1,\vec{r}_1')t_{\alpha_2\downarrow}(x_2,\vec{r}_2')
  ]
  +
  [
  \vec{r}_1'\leftrightarrow{}\vec{r}_2'
  ]
  \,,
\end{align}
and we have made use of the singlet symmetry of the $F$ function.
The cutoff $a$ in the theory specified by
Eqs.~\eqref{eq:effective-hamiltonian} and
\eqref{eq:effective-gamma-general} can be chosen freely, but taking
$a=|\Delta|^{-1}$ is natural as the source term in the original RG
stops contributing at that length scale.

Consider the noninteracting case, $g=1$. There,
\begin{align}
  \label{eq:bose-correlator}
  \frac{
    a^{1-\eta_{2,\alpha\beta}}C_{\alpha\beta}(x\sgn(\tau),|\tau|)
    \sgn(\tau)^{\delta_{\alpha\beta}}
  }{
    a_0^{1-\eta_{2,\alpha\beta}}
    C_{\alpha\beta}(0,a_0)
  }
  =
  \begin{cases}
    a^{-1}[v_F\tau - ix] \,,
    \\
    1 \,,
  \end{cases}
\end{align}
for $\alpha\beta=++$ and $\alpha\beta=+-$, respectively.  One also
notes that
$\Gamma_{-+}\psi_-\psi_+=-\Gamma_{+-}\psi_-\psi_+=\Gamma_{+-}\psi_+\psi_-$,
so we redefine $\Gamma_{+-}\mapsto{}2\Gamma_{+-}$ as the sum of the
two and drop the $-+$ term.  Finally, going into Fourier
representation yields Eqs.~\eqref{eq:gamma-plus-plus}
and~\eqref{eq:gamma-plus-minus-expression}. Due to the integrals
over $x'$ and $\tau'$ extending over the whole range, and the correlation
functions $C_{\alpha\beta}$ being local in frequency and energy,
only certain energies and momenta contribute in the final result.

To find out the effect of interactions, one needs to roughly estimate
the result from Eq.~\eqref{eq:effective-gamma-general}.  First, since
$k_{F,S}^{-1}$ in the superconductor is a short length scale, we take
$F(\vec{r}_1',\tau';\vec{r}_2',0)\mapsto
F(\tau')\delta(\vec{r}_1'-\vec{r}_2')$, $F(\tau')=N_F\Delta
K_0(|\tau|\Delta)$, where $N_F$ is the normal-state DOS at Fermi
energy in the superconductor, and $|\tau|T\ll1$, $T\ll\Delta\ll|M|$.  We
consider tunneling that is local on length scales of $1/\Delta$ (and
$1/T$ and the other low-energy scales) and replace $P_{+-}\sim{\cal
  K}\delta(x')\delta(x-x_1')$ with ${\cal K}$ a constant. Based on the
model in Sec.~\eqref{sec:hgte-rashba-effect}, the Fourier transform of
$P_{++}$ in $x'$ satisfies $P_{++}\sim -i k (z_0/M) P_{+-}$ on long
wavelengths $|k|\ll{}M$, with $P_{+-}$ constant in
$k$. In real space we then have $P_{++}\sim{\cal
  K}\frac{z_0}{M}\partial_{x'}\delta(x')\delta(x-x_1')$.  Within these
assumptions, we get
\begin{subequations}
\label{eq:gamma-interacting-magnitudes}
\begin{align}
  \Gamma_{+-}
  &\sim
  {\cal K}
  N_F
  a^{1-\eta_2}a_0^{2\eta_1-1}
  \Delta
  \int_0^\infty\dd{\tau'}
  K_0(\Delta|\tau'|)
  C_{+-}(\frac{v_F\tau'}{q_0})
  \\
  &=
  {\cal K}N_F(a\Delta)^{1-1/g}(a_0\Delta)^{\frac{g+g^{-1}}{2}-1}
  q(\frac{g^{-1}-g}{2})
  \,,
  \\
  \Gamma_{++}
  &\sim
  {\cal K}
  N_F
  \frac{a^{1-\eta_2}a_0^{2\eta_1-1}
  \Delta z_0}{2|M|}
  \int_0^\infty\dd{\tau'}
  K_0(\Delta|\tau'|)
  \\\notag&\qquad\times
  i\partial_{x'}\Im C_{++}(\frac{v_F\tau'-ix'}{q_0})\rvert_{x'=0}
  \\
  &=
  \frac{i{\cal K}N_F}{2}(a\Delta)^{1-g-1/g}(a_0\Delta)^{\frac{g+g^{-1}}{2}-1}
  \frac{z_0\Delta}{v_F |M|}
  \\\notag&\qquad\times
  q(\frac{g^{-1}+g}{2}-1)
  \,,
\end{align}
\end{subequations}
where $q(x)=2^{x-1}\Gamma(\frac{1+x}{2})^2\sim1$.
The effective tunnel rates obtain an identical scaling in the bare
short-distance cutoff $a_0$ related to interactions, which appears
because of a renormalization of the tunneling elements
$t_{\alpha\sigma}$. The low-energy scaling with $a$ follows the
scaling dimensions in the effective Hamiltonian. Finally,
$\Gamma_{++}$ has an additional factor $z_0\Delta/v_F|M|$ that is a
signature of the Rashba coupling.

The expression \eqref{eq:gamma-plus-plus} for $\Gamma_{++}$ deserves
some comments: First, we know that $F(\omega)\simeq{}1+c\omega^2$ for
$\omega\to0$, so that $\partial_\omega F\rvert_{\omega=0}$ vanishes,
and from Eq.~\eqref{eq:M-symmetry} we know that
$P_{++}(-k)=-P_{++}(k)$ which means that $\partial_k P_{++}$ is even
in $k$ and can be finite at $k=0$.  Note that the gradients
$\partial_\omega$, $\partial_k$ appear because the boson correlation
function $\avg{e^{i\phi_+(x,\tau)}e^{i\phi_+(0,0)}}_0$ vanishes at
$x,\tau\to0$ [see Eq.~\eqref{eq:bose-correlator}], which reflects the
fermionic exclusion principle.  This is the reason why the effective
Hamiltonian contains a term resembling more $\psi_+\psi_+$ than
$\psi_+ k \psi_+$, which is in agreement with the results of
Ref.~\onlinecite{fisher1994-cti}.

Finally, we observe that in the noninteracting case, $\Gamma_{+-}$ is
related to the leading-order off-diagonal Nambu component of the
self-energy.  The factors $\Gamma_{++}$ (and $\Gamma_{+-}$ in the
interacting case) however in general contain additional information,
as the out-integration of short length scales captures the
renormalization from interactions, and the effect of the exclusion
principle when averaging $\psi(x+x')\psi(x)$ over short distances
$x'$.

\section{Current operator}
\label{sec:current-operator}

For completeness, we include here a derivation of
Eq.~\eqref{eq:heisenberg-current-operator} that shows the result
obtained in Ref.~\onlinecite{virtanen2011-dos} applies also to
time-dependent perturbations.  Related results can be found e.g. in
Ref.~\onlinecite{safi1995-tts}, and a special case of the present
result is given in terms of path integrals in
Ref.~\onlinecite{dolcini2005-tpo}.

Consider the Heisenberg equation of motion under a Hamiltonian
$H=H_0+V(t)$, where $H_0$ is given in
Eq.~\eqref{eq:bosonized-hamiltonian}, and the perturbation $V(t)$ is
switched on at $t>0$. Iterating the equation of
motion for $\partial_x\phi$ twice, one obtains
\begin{gather}
  \partial_t^2(\partial_x\phi)
  -
  \partial_x(u^2\partial_x(\partial_x\phi))
  =
  s(x,t)
  \\
  s(x,t) =
  [H_0, [H_0, \partial_x \phi]] - [H, [H, \partial_x\phi]]
  +i[\dot H, \partial_x\phi]
  \,,
\end{gather}
where $\dot H=\dot V$ contains the explicit time dependence of the
Hamiltonian, and $u(x)=v_F/g(x)$ is the renormalized wave velocity.  The
solution to this linear equation can be written in terms of the
retarded Green function of the wave equation on the LHS:
\begin{align}
  \label{eq:current-operator-generic-solution}
  \partial_x\phi(x,t)
  &=
  \partial_x\phi_{\rm 0}(x,t)
  \\\notag
  &+
  \int_{-\infty}^\infty\dd{x'}C^R(x,t;x',0)i[V,\partial_x\phi_{\rm 0}(x',0)]
  \\\notag
  &+
  \int_{0}^\infty\dd{t'}
  \int_{-\infty}^\infty\dd{x'}C^R(x,t;x',t')s(x',t')
  \,,
\end{align}
where $\partial_x\phi_{\rm 0}$ evolves under $H_0$,
and the second term ensures that the initial condition
$\partial_t(\partial_x\phi)=i[H,\partial_x\phi]$ is satisfied
--- this follows from
$\partial_t C^R(x,t;x',t')\rvert_{t\to{}t'+0^+}=\delta(x-x')$.

We can also rewrite $s$ using properties of the fields and $H_0$:
\begin{align}
  s(x,t)
  =
  -\partial_x\bigl(v_Fg(x)^{-2}\frac{\delta V(t)}{\delta\phi(x)}\bigr)
  +
  \partial_t\frac{\delta V(t)}{\delta \vartheta(x)}
  \,,
\end{align}
where we noted the correspondence
\begin{align}
  [A, \partial_x\phi(x)] = -i \frac{\delta A}{\delta \vartheta(x)}
  \,,
  \;
  [A, \partial_x\vartheta(x)] = -i \frac{\delta A}{\delta \phi(x)}
  \,,
\end{align}
valid for functionals $A=A[\vartheta,\phi]$.  The expression for $s$
can be substituted in
Eq.~\eqref{eq:current-operator-generic-solution}, and an integration
by parts transfers the gradients to operate on $C^R$.  One of the
resulting boundary terms cancels the second term in
Eq.~\eqref{eq:current-operator-generic-solution}, and the others
vanish, provided the perturbation $s$ vanishes at $x\to\pm\infty$.

We then find the exact result
\begin{align}
  \label{eq:current-operator-generic-solution-2}
  \partial_x\phi(x,t)
  &=
  \partial_x\phi_{\rm eq}(x,t)
  +
  \int_{0}^\infty\dd{t'}\int_{-\infty}^\infty\dd{x'}\Bigl[
  \\\notag&\quad
  v_Fg(x')^{-2}[\partial_{x'}C^R(x,t;x',t')]\frac{\delta V(t')}{\delta\phi(x')}
  \\\notag&\quad
  -
  [\partial_{t'} C^R(x,t;x',t')] \frac{\delta V(t')}{\delta\vartheta(x')}
  \Bigr]
  \,.
\end{align}
We can simplify this further by making use of properties of the 1-D
wave equation. The Green function $C^R(x,t;x',t')$ satisfies the
initial value problem
\begin{gather}
  [\partial_t^2 - \partial_x u^2\partial_x]C^R = 0
  \,,\;\text{at $t>t'$}\,,
  \\
  C^R = 0 \,,\quad
  \partial_t C^R = \delta(x-x') \,,
  \;\text{at $t=t'$.}
\end{gather}
If $u$ is a constant, the solution is a sum of two wavefronts
$C^R=C_+^R + C_-^R$,
$C_\pm^R=(4u)^{-1}\theta(t-t')\sgn[\pm(x'-x)+u(t-t')]$.  This is
valid in the limit $t\to{}t'$ also if $u$ is smoothly spatially
varying --- the wave equation only sees $u$ around $x'$. Since the wave
equation is linear and its solution is unique, the Green function can
always be decomposed to these two parts.  Let us now define $D_\pm =
2 \partial_t C^R_{\pm}$ and $F_\pm = 2\partial_{x'} C^R_{\pm}$. They
satisfy the wave equation at $t>t'$, and the initial conditions are
inherited from the $t\to{}t'$ behavior of $C^R_{\pm}$:
\begin{gather}
  D_\pm = \delta(x-x') \,,\;
  \partial_t D_\pm = \mp u(x')\delta'(x-x')
  \,,
  \\
  F_\pm = \pm u(x')^{-1}\delta(x-x') \,,\;
  \partial_t F_\pm = -\delta'(x-x') \,.
\end{gather}
Due to linearity, clearly $F_\pm = \pm u(x') D_\pm$.  Because
$C^R(x,t;x',t')=C^R(x,x',t-t')$, we then find
$\partial_{t'}C^R=-\partial_tC^R=-\frac{D_+ + D_-}{2}$ and
$\partial_{x'}C^R=\frac{F_++F_-}{2}=u(x')\frac{D_+ - D_-}{2}$.
Substituting these to
Eq.~\eqref{eq:current-operator-generic-solution-2} and defining
$j_\pm=\frac{1}{2\sqrt{\pi}}[\frac{\delta V}{\delta\phi}\pm\frac{\delta
  V}{\delta\vartheta}]\equiv\frac{\delta V}{\delta \phi_\pm}$,
we arrive at Eq.~\eqref{eq:heisenberg-current-operator}.


\begin{thebibliography}{44}%
\makeatletter
\providecommand \@ifxundefined [1]{%
 \@ifx{#1\undefined}
}%
\providecommand \@ifnum [1]{%
 \ifnum #1\expandafter \@firstoftwo
 \else \expandafter \@secondoftwo
 \fi
}%
\providecommand \@ifx [1]{%
 \ifx #1\expandafter \@firstoftwo
 \else \expandafter \@secondoftwo
 \fi
}%
\providecommand \natexlab [1]{#1}%
\providecommand \enquote  [1]{``#1''}%
\providecommand \bibnamefont  [1]{#1}%
\providecommand \bibfnamefont [1]{#1}%
\providecommand \citenamefont [1]{#1}%
\providecommand \href@noop [0]{\@secondoftwo}%
\providecommand \href [0]{\begingroup \@sanitize@url \@href}%
\providecommand \@href[1]{\@@startlink{#1}\@@href}%
\providecommand \@@href[1]{\endgroup#1\@@endlink}%
\providecommand \@sanitize@url [0]{\catcode `\\12\catcode `\$12\catcode
  `\&12\catcode `\#12\catcode `\^12\catcode `\_12\catcode `\%12\relax}%
\providecommand \@@startlink[1]{}%
\providecommand \@@endlink[0]{}%
\providecommand \url  [0]{\begingroup\@sanitize@url \@url }%
\providecommand \@url [1]{\endgroup\@href {#1}{\urlprefix }}%
\providecommand \urlprefix  [0]{URL }%
\providecommand \Eprint [0]{\href }%
\providecommand \doibase [0]{http://dx.doi.org/}%
\providecommand \selectlanguage [0]{\@gobble}%
\providecommand \bibinfo  [0]{\@secondoftwo}%
\providecommand \bibfield  [0]{\@secondoftwo}%
\providecommand \translation [1]{[#1]}%
\providecommand \BibitemOpen [0]{}%
\providecommand \bibitemStop [0]{}%
\providecommand \bibitemNoStop [0]{.\EOS\space}%
\providecommand \EOS [0]{\spacefactor3000\relax}%
\providecommand \BibitemShut  [1]{\csname bibitem#1\endcsname}%
\let\auto@bib@innerbib\@empty
\bibitem [{\citenamefont {Kane}\ and\ \citenamefont
  {Mele}(2005)}]{kane2005-qsh}%
  \BibitemOpen
  \bibfield  {author} {\bibinfo {author} {\bibfnamefont {C.~L.}\ \bibnamefont
  {Kane}}\ and\ \bibinfo {author} {\bibfnamefont {E.~J.}\ \bibnamefont
  {Mele}},\ }\href {\doibase 10.1103/PhysRevLett.95.226801} {\bibfield
  {journal} {\bibinfo  {journal} {Phys. Rev. Lett.}\ }\textbf {\bibinfo
  {volume} {95}},\ \bibinfo {pages} {226801} (\bibinfo {year}
  {2005})}\BibitemShut {NoStop}%
\bibitem [{\citenamefont {Bernevig}\ \emph {et~al.}(2006)\citenamefont
  {Bernevig}, \citenamefont {Hughes},\ and\ \citenamefont
  {Zhang}}]{bernevig2006-qsh}%
  \BibitemOpen
  \bibfield  {author} {\bibinfo {author} {\bibfnamefont {B.~A.}\ \bibnamefont
  {Bernevig}}, \bibinfo {author} {\bibfnamefont {T.~L.}\ \bibnamefont
  {Hughes}}, \ and\ \bibinfo {author} {\bibfnamefont {S.~C.}\ \bibnamefont
  {Zhang}},\ }\href {\doibase 10.1126/science.1133734} {\bibfield  {journal}
  {\bibinfo  {journal} {Science}\ }\textbf {\bibinfo {volume} {314}},\ \bibinfo
  {pages} {1757} (\bibinfo {year} {2006})}\BibitemShut {NoStop}%
\bibitem [{\citenamefont {Wu}\ \emph {et~al.}(2006)\citenamefont {Wu},
  \citenamefont {Bernevig},\ and\ \citenamefont {Zhang}}]{wu2006-hla}%
  \BibitemOpen
  \bibfield  {author} {\bibinfo {author} {\bibfnamefont {C.}~\bibnamefont
  {Wu}}, \bibinfo {author} {\bibfnamefont {B.~A.}\ \bibnamefont {Bernevig}}, \
  and\ \bibinfo {author} {\bibfnamefont {S.-C.}\ \bibnamefont {Zhang}},\ }\href
  {\doibase 10.1103/PhysRevLett.96.106401} {\bibfield  {journal} {\bibinfo
  {journal} {Phys. Rev. Lett.}\ }\textbf {\bibinfo {volume} {96}},\ \bibinfo
  {pages} {106401} (\bibinfo {year} {2006})}\BibitemShut {NoStop}%
\bibitem [{\citenamefont {Xu}\ and\ \citenamefont {Moore}(2006)}]{xu2006-soq}%
  \BibitemOpen
  \bibfield  {author} {\bibinfo {author} {\bibfnamefont {C.}~\bibnamefont
  {Xu}}\ and\ \bibinfo {author} {\bibfnamefont {J.~E.}\ \bibnamefont {Moore}},\
  }\href {\doibase 10.1103/PhysRevB.73.045322} {\bibfield  {journal} {\bibinfo
  {journal} {Phys. Rev. B}\ }\textbf {\bibinfo {volume} {73}},\ \bibinfo
  {pages} {045322} (\bibinfo {year} {2006})}\BibitemShut {NoStop}%
\bibitem [{\citenamefont {K\"onig}\ \emph {et~al.}(2007)\citenamefont
  {K\"onig}, \citenamefont {Wiedmann}, \citenamefont {Br\"une}, \citenamefont
  {Roth}, \citenamefont {Buhmann}, \citenamefont {Molenkamp}, \citenamefont
  {Qi},\ and\ \citenamefont {Zhang}}]{konig2007-qsh}%
  \BibitemOpen
  \bibfield  {author} {\bibinfo {author} {\bibfnamefont {M.}~\bibnamefont
  {K\"onig}}, \bibinfo {author} {\bibfnamefont {S.}~\bibnamefont {Wiedmann}},
  \bibinfo {author} {\bibfnamefont {C.}~\bibnamefont {Br\"une}}, \bibinfo
  {author} {\bibfnamefont {A.}~\bibnamefont {Roth}}, \bibinfo {author}
  {\bibfnamefont {H.}~\bibnamefont {Buhmann}}, \bibinfo {author} {\bibfnamefont
  {L.~W.}\ \bibnamefont {Molenkamp}}, \bibinfo {author} {\bibfnamefont {X.-L.}\
  \bibnamefont {Qi}}, \ and\ \bibinfo {author} {\bibfnamefont {S.-C.}\
  \bibnamefont {Zhang}},\ }\href {\doibase 10.1126/science.1148047} {\bibfield
  {journal} {\bibinfo  {journal} {Science}\ }\textbf {\bibinfo {volume}
  {318}},\ \bibinfo {pages} {766} (\bibinfo {year} {2007})}\BibitemShut
  {NoStop}%
\bibitem [{\citenamefont {Roth}\ \emph {et~al.}(2009)\citenamefont {Roth},
  \citenamefont {Br\"une}, \citenamefont {Buhmann}, \citenamefont {Molenkamp},
  \citenamefont {Maciejko}, \citenamefont {Qi},\ and\ \citenamefont
  {Zhang}}]{roth2009-nti}%
  \BibitemOpen
  \bibfield  {author} {\bibinfo {author} {\bibfnamefont {A.}~\bibnamefont
  {Roth}}, \bibinfo {author} {\bibfnamefont {C.}~\bibnamefont {Br\"une}},
  \bibinfo {author} {\bibfnamefont {H.}~\bibnamefont {Buhmann}}, \bibinfo
  {author} {\bibfnamefont {L.~W.}\ \bibnamefont {Molenkamp}}, \bibinfo {author}
  {\bibfnamefont {J.}~\bibnamefont {Maciejko}}, \bibinfo {author}
  {\bibfnamefont {X.-L.}\ \bibnamefont {Qi}}, \ and\ \bibinfo {author}
  {\bibfnamefont {S.-C.}\ \bibnamefont {Zhang}},\ }\href {\doibase
  10.1126/science.1174736} {\bibfield  {journal} {\bibinfo  {journal}
  {Science}\ }\textbf {\bibinfo {volume} {325}},\ \bibinfo {pages} {294}
  (\bibinfo {year} {2009})}\BibitemShut {NoStop}%
\bibitem [{\citenamefont {Qi}\ and\ \citenamefont {Zhang}(2011)}]{qi2011-tia}%
  \BibitemOpen
  \bibfield  {author} {\bibinfo {author} {\bibfnamefont {X.-L.}\ \bibnamefont
  {Qi}}\ and\ \bibinfo {author} {\bibfnamefont {S.-C.}\ \bibnamefont {Zhang}},\
  }\href {\doibase 10.1103/RevModPhys.83.1057} {\bibfield  {journal} {\bibinfo
  {journal} {Rev. Mod. Phys.}\ }\textbf {\bibinfo {volume} {83}},\ \bibinfo
  {pages} {1057} (\bibinfo {year} {2011})}\BibitemShut {NoStop}%
\bibitem [{\citenamefont {Hasan}\ and\ \citenamefont
  {Kane}(2010)}]{hasan2010-cti}%
  \BibitemOpen
  \bibfield  {author} {\bibinfo {author} {\bibfnamefont {M.~Z.}\ \bibnamefont
  {Hasan}}\ and\ \bibinfo {author} {\bibfnamefont {C.~L.}\ \bibnamefont
  {Kane}},\ }\href {\doibase 10.1103/RevModPhys.82.3045} {\bibfield  {journal}
  {\bibinfo  {journal} {Rev. Mod. Phys.}\ }\textbf {\bibinfo {volume} {82}},\
  \bibinfo {pages} {3045} (\bibinfo {year} {2010})}\BibitemShut {NoStop}%
\bibitem [{\citenamefont {Fu}\ and\ \citenamefont {Kane}(2008)}]{fu2008-spe}%
  \BibitemOpen
  \bibfield  {author} {\bibinfo {author} {\bibfnamefont {L.}~\bibnamefont
  {Fu}}\ and\ \bibinfo {author} {\bibfnamefont {C.~L.}\ \bibnamefont {Kane}},\
  }\href {\doibase 10.1103/PhysRevLett.100.096407} {\bibfield  {journal}
  {\bibinfo  {journal} {Phys. Rev. Lett.}\ }\textbf {\bibinfo {volume} {100}},\
  \bibinfo {pages} {096407} (\bibinfo {year} {2008})}\BibitemShut {NoStop}%
\bibitem [{\citenamefont {Linder}\ \emph {et~al.}(2010)\citenamefont {Linder},
  \citenamefont {Tanaka}, \citenamefont {Yokoyama}, \citenamefont {Sudb\o{}},\
  and\ \citenamefont {Nagaosa}}]{linder2010-ust}%
  \BibitemOpen
  \bibfield  {author} {\bibinfo {author} {\bibfnamefont {J.}~\bibnamefont
  {Linder}}, \bibinfo {author} {\bibfnamefont {Y.}~\bibnamefont {Tanaka}},
  \bibinfo {author} {\bibfnamefont {T.}~\bibnamefont {Yokoyama}}, \bibinfo
  {author} {\bibfnamefont {A.}~\bibnamefont {Sudb\o{}}}, \ and\ \bibinfo
  {author} {\bibfnamefont {N.}~\bibnamefont {Nagaosa}},\ }\href {\doibase
  10.1103/PhysRevLett.104.067001} {\bibfield  {journal} {\bibinfo  {journal}
  {Phys. Rev. Lett.}\ }\textbf {\bibinfo {volume} {104}},\ \bibinfo {pages}
  {067001} (\bibinfo {year} {2010})}\BibitemShut {NoStop}%
\bibitem [{\citenamefont {Stanescu}\ \emph {et~al.}(2010)\citenamefont
  {Stanescu}, \citenamefont {Sau}, \citenamefont {Lutchyn},\ and\ \citenamefont
  {Das~Sarma}}]{stanescu2010-pea}%
  \BibitemOpen
  \bibfield  {author} {\bibinfo {author} {\bibfnamefont {T.~D.}\ \bibnamefont
  {Stanescu}}, \bibinfo {author} {\bibfnamefont {J.~D.}\ \bibnamefont {Sau}},
  \bibinfo {author} {\bibfnamefont {R.~M.}\ \bibnamefont {Lutchyn}}, \ and\
  \bibinfo {author} {\bibfnamefont {S.}~\bibnamefont {Das~Sarma}},\ }\href
  {\doibase 10.1103/PhysRevB.81.241310} {\bibfield  {journal} {\bibinfo
  {journal} {Phys. Rev. B}\ }\textbf {\bibinfo {volume} {81}},\ \bibinfo
  {pages} {241310} (\bibinfo {year} {2010})}\BibitemShut {NoStop}%
\bibitem [{\citenamefont {Sau}\ \emph {et~al.}(2010{\natexlab{a}})\citenamefont
  {Sau}, \citenamefont {Lutchyn}, \citenamefont {Tewari},\ and\ \citenamefont
  {Das~Sarma}}]{sau2010-gnp}%
  \BibitemOpen
  \bibfield  {author} {\bibinfo {author} {\bibfnamefont {J.~D.}\ \bibnamefont
  {Sau}}, \bibinfo {author} {\bibfnamefont {R.~M.}\ \bibnamefont {Lutchyn}},
  \bibinfo {author} {\bibfnamefont {S.}~\bibnamefont {Tewari}}, \ and\ \bibinfo
  {author} {\bibfnamefont {S.}~\bibnamefont {Das~Sarma}},\ }\href {\doibase
  10.1103/PhysRevLett.104.040502} {\bibfield  {journal} {\bibinfo  {journal}
  {Phys. Rev. Lett.}\ }\textbf {\bibinfo {volume} {104}},\ \bibinfo {pages}
  {040502} (\bibinfo {year} {2010}{\natexlab{a}})}\BibitemShut {NoStop}%
\bibitem [{\citenamefont {Sau}\ \emph {et~al.}(2010{\natexlab{b}})\citenamefont
  {Sau}, \citenamefont {Lutchyn}, \citenamefont {Tewari},\ and\ \citenamefont
  {Das~Sarma}}]{sau2010-rom}%
  \BibitemOpen
  \bibfield  {author} {\bibinfo {author} {\bibfnamefont {J.~D.}\ \bibnamefont
  {Sau}}, \bibinfo {author} {\bibfnamefont {R.~M.}\ \bibnamefont {Lutchyn}},
  \bibinfo {author} {\bibfnamefont {S.}~\bibnamefont {Tewari}}, \ and\ \bibinfo
  {author} {\bibfnamefont {S.}~\bibnamefont {Das~Sarma}},\ }\href {\doibase
  10.1103/PhysRevB.82.094522} {\bibfield  {journal} {\bibinfo  {journal} {Phys.
  Rev. B}\ }\textbf {\bibinfo {volume} {82}},\ \bibinfo {pages} {094522}
  (\bibinfo {year} {2010}{\natexlab{b}})}\BibitemShut {NoStop}%
\bibitem [{\citenamefont {Alicea}(2010)}]{alicea2010-mfi}%
  \BibitemOpen
  \bibfield  {author} {\bibinfo {author} {\bibfnamefont {J.}~\bibnamefont
  {Alicea}},\ }\href {\doibase 10.1103/PhysRevB.81.125318} {\bibfield
  {journal} {\bibinfo  {journal} {Phys. Rev. B}\ }\textbf {\bibinfo {volume}
  {81}},\ \bibinfo {pages} {125318} (\bibinfo {year} {2010})}\BibitemShut
  {NoStop}%
\bibitem [{\citenamefont {Linder}\ and\ \citenamefont
  {Sudb\o{}}(2010)}]{linder2010-mfm}%
  \BibitemOpen
  \bibfield  {author} {\bibinfo {author} {\bibfnamefont {J.}~\bibnamefont
  {Linder}}\ and\ \bibinfo {author} {\bibfnamefont {A.}~\bibnamefont
  {Sudb\o{}}},\ }\href {\doibase 10.1103/PhysRevB.82.085314} {\bibfield
  {journal} {\bibinfo  {journal} {Phys. Rev. B}\ }\textbf {\bibinfo {volume}
  {82}},\ \bibinfo {pages} {085314} (\bibinfo {year} {2010})}\BibitemShut
  {NoStop}%
\bibitem [{\citenamefont {Fu}\ and\ \citenamefont {Kane}(2009)}]{fu2009-jca}%
  \BibitemOpen
  \bibfield  {author} {\bibinfo {author} {\bibfnamefont {L.}~\bibnamefont
  {Fu}}\ and\ \bibinfo {author} {\bibfnamefont {C.~L.}\ \bibnamefont {Kane}},\
  }\href {\doibase 10.1103/PhysRevB.79.161408} {\bibfield  {journal} {\bibinfo
  {journal} {Phys. Rev. B}\ }\textbf {\bibinfo {volume} {79}},\ \bibinfo
  {pages} {161408} (\bibinfo {year} {2009})}\BibitemShut {NoStop}%
\bibitem [{\citenamefont {Sato}\ \emph {et~al.}(2010)\citenamefont {Sato},
  \citenamefont {Loss},\ and\ \citenamefont {Tserkovnyak}}]{sato2010-cii}%
  \BibitemOpen
  \bibfield  {author} {\bibinfo {author} {\bibfnamefont {K.}~\bibnamefont
  {Sato}}, \bibinfo {author} {\bibfnamefont {D.}~\bibnamefont {Loss}}, \ and\
  \bibinfo {author} {\bibfnamefont {Y.}~\bibnamefont {Tserkovnyak}},\ }\href
  {\doibase 10.1103/PhysRevLett.105.226401} {\bibfield  {journal} {\bibinfo
  {journal} {Phys. Rev. Lett.}\ }\textbf {\bibinfo {volume} {105}},\ \bibinfo
  {pages} {226401} (\bibinfo {year} {2010})}\BibitemShut {NoStop}%
\bibitem [{\citenamefont {Adroguer}\ \emph {et~al.}(2010)\citenamefont
  {Adroguer}, \citenamefont {Grenier}, \citenamefont {Carpentier},
  \citenamefont {Cayssol}, \citenamefont {Degiovanni},\ and\ \citenamefont
  {Orignac}}]{adroguer2010-phe}%
  \BibitemOpen
  \bibfield  {author} {\bibinfo {author} {\bibfnamefont {P.}~\bibnamefont
  {Adroguer}}, \bibinfo {author} {\bibfnamefont {C.}~\bibnamefont {Grenier}},
  \bibinfo {author} {\bibfnamefont {D.}~\bibnamefont {Carpentier}}, \bibinfo
  {author} {\bibfnamefont {J.}~\bibnamefont {Cayssol}}, \bibinfo {author}
  {\bibfnamefont {P.}~\bibnamefont {Degiovanni}}, \ and\ \bibinfo {author}
  {\bibfnamefont {E.}~\bibnamefont {Orignac}},\ }\href {\doibase
  10.1103/PhysRevB.82.081303} {\bibfield  {journal} {\bibinfo  {journal} {Phys.
  Rev. B}\ }\textbf {\bibinfo {volume} {82}},\ \bibinfo {pages} {081303}
  (\bibinfo {year} {2010})}\BibitemShut {NoStop}%
\bibitem [{\citenamefont {Black-Schaffer}(2011)}]{black-schaffer2011-ssp}%
  \BibitemOpen
  \bibfield  {author} {\bibinfo {author} {\bibfnamefont {A.~M.}\ \bibnamefont
  {Black-Schaffer}},\ }\href {\doibase 10.1103/PhysRevB.83.060504} {\bibfield
  {journal} {\bibinfo  {journal} {Phys. Rev. B}\ }\textbf {\bibinfo {volume}
  {83}},\ \bibinfo {pages} {060504} (\bibinfo {year} {2011})}\BibitemShut
  {NoStop}%
\bibitem [{\citenamefont {Fisher}(1994)}]{fisher1994-cti}%
  \BibitemOpen
  \bibfield  {author} {\bibinfo {author} {\bibfnamefont {M.~P.~A.}\
  \bibnamefont {Fisher}},\ }\href {\doibase 10.1103/PhysRevB.49.14550}
  {\bibfield  {journal} {\bibinfo  {journal} {Phys. Rev. B}\ }\textbf {\bibinfo
  {volume} {49}},\ \bibinfo {pages} {14550} (\bibinfo {year}
  {1994})}\BibitemShut {NoStop}%
\bibitem [{\citenamefont {Fazio}\ \emph {et~al.}(1996)\citenamefont {Fazio},
  \citenamefont {Hekking},\ and\ \citenamefont {Odintsov}}]{fazio1996-daa}%
  \BibitemOpen
  \bibfield  {author} {\bibinfo {author} {\bibfnamefont {R.}~\bibnamefont
  {Fazio}}, \bibinfo {author} {\bibfnamefont {F.~W.~J.}\ \bibnamefont
  {Hekking}}, \ and\ \bibinfo {author} {\bibfnamefont {A.~A.}\ \bibnamefont
  {Odintsov}},\ }\href {\doibase 10.1103/PhysRevB.53.6653} {\bibfield
  {journal} {\bibinfo  {journal} {Phys. Rev. B}\ }\textbf {\bibinfo {volume}
  {53}},\ \bibinfo {pages} {6653} (\bibinfo {year} {1996})}\BibitemShut
  {NoStop}%
\bibitem [{\citenamefont {Fazio}\ \emph {et~al.}(1995)\citenamefont {Fazio},
  \citenamefont {Hekking},\ and\ \citenamefont {Odintsov}}]{fazio1995-jct}%
  \BibitemOpen
  \bibfield  {author} {\bibinfo {author} {\bibfnamefont {R.}~\bibnamefont
  {Fazio}}, \bibinfo {author} {\bibfnamefont {F.~W.~J.}\ \bibnamefont
  {Hekking}}, \ and\ \bibinfo {author} {\bibfnamefont {A.~A.}\ \bibnamefont
  {Odintsov}},\ }\href {\doibase 10.1103/PhysRevLett.74.1843} {\bibfield
  {journal} {\bibinfo  {journal} {Phys. Rev. Lett.}\ }\textbf {\bibinfo
  {volume} {74}},\ \bibinfo {pages} {1843} (\bibinfo {year}
  {1995})}\BibitemShut {NoStop}%
\bibitem [{\citenamefont {Pugnetti}\ \emph {et~al.}(2007)\citenamefont
  {Pugnetti}, \citenamefont {Dolcini},\ and\ \citenamefont
  {Fazio}}]{pugnetti2007-dje}%
  \BibitemOpen
  \bibfield  {author} {\bibinfo {author} {\bibfnamefont {S.}~\bibnamefont
  {Pugnetti}}, \bibinfo {author} {\bibfnamefont {F.}~\bibnamefont {Dolcini}}, \
  and\ \bibinfo {author} {\bibfnamefont {R.}~\bibnamefont {Fazio}},\
  }\href@noop {} {\bibfield  {journal} {\bibinfo  {journal} {Solid State
  Commun.}\ }\textbf {\bibinfo {volume} {144}},\ \bibinfo {pages} {551}
  (\bibinfo {year} {2007})}\BibitemShut {NoStop}%
\bibitem [{\citenamefont {Novik}\ \emph {et~al.}(2005)\citenamefont {Novik},
  \citenamefont {Pfeuffer-Jeschke}, \citenamefont {Jungwirth}, \citenamefont
  {Latussek}, \citenamefont {Becker}, \citenamefont {Landwehr}, \citenamefont
  {Buhmann},\ and\ \citenamefont {Molenkamp}}]{novik2005-bso}%
  \BibitemOpen
  \bibfield  {author} {\bibinfo {author} {\bibfnamefont {E.~G.}\ \bibnamefont
  {Novik}}, \bibinfo {author} {\bibfnamefont {A.}~\bibnamefont
  {Pfeuffer-Jeschke}}, \bibinfo {author} {\bibfnamefont {T.}~\bibnamefont
  {Jungwirth}}, \bibinfo {author} {\bibfnamefont {V.}~\bibnamefont {Latussek}},
  \bibinfo {author} {\bibfnamefont {C.~R.}\ \bibnamefont {Becker}}, \bibinfo
  {author} {\bibfnamefont {G.}~\bibnamefont {Landwehr}}, \bibinfo {author}
  {\bibfnamefont {H.}~\bibnamefont {Buhmann}}, \ and\ \bibinfo {author}
  {\bibfnamefont {L.~W.}\ \bibnamefont {Molenkamp}},\ }\href {\doibase
  10.1103/PhysRevB.72.035321} {\bibfield  {journal} {\bibinfo  {journal} {Phys.
  Rev. B}\ }\textbf {\bibinfo {volume} {72}},\ \bibinfo {pages} {035321}
  (\bibinfo {year} {2005})}\BibitemShut {NoStop}%
\bibitem [{\citenamefont {Zhou}\ \emph {et~al.}(2008)\citenamefont {Zhou},
  \citenamefont {Lu}, \citenamefont {Chu}, \citenamefont {Shen},\ and\
  \citenamefont {Niu}}]{zhou2008-fse}%
  \BibitemOpen
  \bibfield  {author} {\bibinfo {author} {\bibfnamefont {B.}~\bibnamefont
  {Zhou}}, \bibinfo {author} {\bibfnamefont {H.-Z.}\ \bibnamefont {Lu}},
  \bibinfo {author} {\bibfnamefont {R.-L.}\ \bibnamefont {Chu}}, \bibinfo
  {author} {\bibfnamefont {S.-Q.}\ \bibnamefont {Shen}}, \ and\ \bibinfo
  {author} {\bibfnamefont {Q.}~\bibnamefont {Niu}},\ }\href {\doibase
  10.1103/PhysRevLett.101.246807} {\bibfield  {journal} {\bibinfo  {journal}
  {Phys. Rev. Lett.}\ }\textbf {\bibinfo {volume} {101}},\ \bibinfo {pages}
  {246807} (\bibinfo {year} {2008})}\BibitemShut {NoStop}%
\bibitem [{\citenamefont {Abrikosov}\ \emph {et~al.}(1975)\citenamefont
  {Abrikosov}, \citenamefont {Gorkov},\ and\ \citenamefont
  {Dzyaloshinski}}]{abrikosov1975-moq}%
  \BibitemOpen
  \bibfield  {author} {\bibinfo {author} {\bibfnamefont {A.~A.}\ \bibnamefont
  {Abrikosov}}, \bibinfo {author} {\bibfnamefont {L.~P.}\ \bibnamefont
  {Gorkov}}, \ and\ \bibinfo {author} {\bibfnamefont {I.~E.}\ \bibnamefont
  {Dzyaloshinski}},\ }\href@noop {} {\emph {\bibinfo {title} {Methods of
  quantum field theory in statistical physics}}}\ (\bibinfo  {publisher} {Dover
  publications, Inc.},\ \bibinfo {address} {New York},\ \bibinfo {year}
  {1975})\BibitemShut {NoStop}%
\bibitem [{\citenamefont {Rammer}\ and\ \citenamefont
  {Smith}(1986)}]{rammer86}%
  \BibitemOpen
  \bibfield  {author} {\bibinfo {author} {\bibfnamefont {J.}~\bibnamefont
  {Rammer}}\ and\ \bibinfo {author} {\bibfnamefont {H.}~\bibnamefont {Smith}},\
  }\href@noop {} {\bibfield  {journal} {\bibinfo  {journal} {Rev. Mod. Phys.}\
  }\textbf {\bibinfo {volume} {58}},\ \bibinfo {pages} {323} (\bibinfo {year}
  {1986})}\BibitemShut {NoStop}%
\bibitem [{\citenamefont {Rothe}\ \emph {et~al.}(2010)\citenamefont {Rothe},
  \citenamefont {Reinthaler}, \citenamefont {Liu}, \citenamefont {Molenkamp},
  \citenamefont {Zhang},\ and\ \citenamefont {Hankiewicz}}]{rothe2010-fod}%
  \BibitemOpen
  \bibfield  {author} {\bibinfo {author} {\bibfnamefont {D.~G.}\ \bibnamefont
  {Rothe}}, \bibinfo {author} {\bibfnamefont {R.~W.}\ \bibnamefont
  {Reinthaler}}, \bibinfo {author} {\bibfnamefont {C.-X.}\ \bibnamefont {Liu}},
  \bibinfo {author} {\bibfnamefont {L.~W.}\ \bibnamefont {Molenkamp}}, \bibinfo
  {author} {\bibfnamefont {S.-C.}\ \bibnamefont {Zhang}}, \ and\ \bibinfo
  {author} {\bibfnamefont {E.~M.}\ \bibnamefont {Hankiewicz}},\ }\href@noop {}
  {\bibfield  {journal} {\bibinfo  {journal} {New Journal of Physics}\ }\textbf
  {\bibinfo {volume} {12}},\ \bibinfo {pages} {065012} (\bibinfo {year}
  {2010})}\BibitemShut {NoStop}%
\bibitem [{\citenamefont {K\"onig}\ \emph {et~al.}(2008)\citenamefont
  {K\"onig}, \citenamefont {Buhmann}, \citenamefont {Molenkamp}, \citenamefont
  {Hughes}, \citenamefont {C.-X.}, \citenamefont {X.-L.},\ and\ \citenamefont
  {Zhang}}]{konig2008-qsh}%
  \BibitemOpen
  \bibfield  {author} {\bibinfo {author} {\bibfnamefont {M.}~\bibnamefont
  {K\"onig}}, \bibinfo {author} {\bibfnamefont {H.}~\bibnamefont {Buhmann}},
  \bibinfo {author} {\bibfnamefont {L.~W.}\ \bibnamefont {Molenkamp}}, \bibinfo
  {author} {\bibfnamefont {T.}~\bibnamefont {Hughes}}, \bibinfo {author}
  {\bibfnamefont {L.}~\bibnamefont {C.-X.}}, \bibinfo {author} {\bibfnamefont
  {Q.}~\bibnamefont {X.-L.}}, \ and\ \bibinfo {author} {\bibfnamefont {S.-C.}\
  \bibnamefont {Zhang}},\ }\href {\doibase 10.1143/JPSJ.77.031007} {\bibfield
  {journal} {\bibinfo  {journal} {J. Phys. Soc. Japan}\ }\textbf {\bibinfo
  {volume} {77}},\ \bibinfo {pages} {031007} (\bibinfo {year}
  {2008})}\BibitemShut {NoStop}%
\bibitem [{Note1()}]{Note1}%
  \BibitemOpen
  \bibinfo {note} {Because the analytical edge state wave functions have a
  discontinuous derivative due to the boundary condition, the matrix element
  $\DOTSI \intop \ilimits@ \protect \mathrm {d}y\protect \tmspace +\thinmuskip
  {.1667em}\phi (y)^*\partial _y^3\psi (y)$ is better rewritten as $\protect
  \frac {1}{2}\DOTSI \intop \ilimits@ \protect \mathrm {d}y\protect \tmspace
  +\thinmuskip {.1667em}[\partial _y^2\phi (y)^*\partial _y\psi (y)-\partial
  _y\phi (y)^*\partial _y^2\psi (y)]$, to remove the need to evaluate boundary
  terms.}\BibitemShut {Stop}%
\bibitem [{\citenamefont {Giamarchi}(2004)}]{giamarchi2004-qpi}%
  \BibitemOpen
  \bibfield  {author} {\bibinfo {author} {\bibfnamefont {T.}~\bibnamefont
  {Giamarchi}},\ }\href@noop {} {\emph {\bibinfo {title} {Quantum physics in
  one dimension}}}\ (\bibinfo  {publisher} {Oxford University Press},\ \bibinfo
  {year} {2004})\BibitemShut {NoStop}%
\bibitem [{\citenamefont {Virtanen}\ and\ \citenamefont
  {Recher}(2011)}]{virtanen2011-dos}%
  \BibitemOpen
  \bibfield  {author} {\bibinfo {author} {\bibfnamefont {P.}~\bibnamefont
  {Virtanen}}\ and\ \bibinfo {author} {\bibfnamefont {P.}~\bibnamefont
  {Recher}},\ }\href {\doibase 10.1103/PhysRevB.83.115332} {\bibfield
  {journal} {\bibinfo  {journal} {Phys. Rev. B}\ }\textbf {\bibinfo {volume}
  {83}},\ \bibinfo {pages} {115332} (\bibinfo {year} {2011})}\BibitemShut
  {NoStop}%
\bibitem [{Note2()}]{Note2}%
  \BibitemOpen
  \bibinfo {note} {This can be checked using the Bogoliubov--de Gennes
  equation.}\BibitemShut {Stop}%
\bibitem [{\citenamefont {Pothier}\ \emph {et~al.}(1994)\citenamefont
  {Pothier}, \citenamefont {Gu\'eron}, \citenamefont {Esteve},\ and\
  \citenamefont {Devoret}}]{pothier1994-fac}%
  \BibitemOpen
  \bibfield  {author} {\bibinfo {author} {\bibfnamefont {H.}~\bibnamefont
  {Pothier}}, \bibinfo {author} {\bibfnamefont {S.}~\bibnamefont {Gu\'eron}},
  \bibinfo {author} {\bibfnamefont {D.}~\bibnamefont {Esteve}}, \ and\ \bibinfo
  {author} {\bibfnamefont {M.~H.}\ \bibnamefont {Devoret}},\ }\href {\doibase
  10.1103/PhysRevLett.73.2488} {\bibfield  {journal} {\bibinfo  {journal}
  {Phys. Rev. Lett.}\ }\textbf {\bibinfo {volume} {73}},\ \bibinfo {pages}
  {2488} (\bibinfo {year} {1994})}\BibitemShut {NoStop}%
\bibitem [{\citenamefont {Hekking}\ and\ \citenamefont
  {Nazarov}(1993)}]{hekking1993-iot}%
  \BibitemOpen
  \bibfield  {author} {\bibinfo {author} {\bibfnamefont {F.~W.~J.}\
  \bibnamefont {Hekking}}\ and\ \bibinfo {author} {\bibfnamefont {Y.~V.}\
  \bibnamefont {Nazarov}},\ }\href {\doibase 10.1103/PhysRevLett.71.1625}
  {\bibfield  {journal} {\bibinfo  {journal} {Phys. Rev. Lett.}\ }\textbf
  {\bibinfo {volume} {71}},\ \bibinfo {pages} {1625} (\bibinfo {year}
  {1993})}\BibitemShut {NoStop}%
\bibitem [{\citenamefont {Volkov}\ and\ \citenamefont
  {Takayanagi}(1997)}]{volkov1997-lpe}%
  \BibitemOpen
  \bibfield  {author} {\bibinfo {author} {\bibfnamefont {A.~F.}\ \bibnamefont
  {Volkov}}\ and\ \bibinfo {author} {\bibfnamefont {H.}~\bibnamefont
  {Takayanagi}},\ }\href {\doibase 10.1103/PhysRevB.56.11184} {\bibfield
  {journal} {\bibinfo  {journal} {Phys. Rev. B}\ }\textbf {\bibinfo {volume}
  {56}},\ \bibinfo {pages} {11184} (\bibinfo {year} {1997})}\BibitemShut
  {NoStop}%
\bibitem [{\citenamefont {Egger}\ and\ \citenamefont
  {Grabert}(1998)}]{egger1998-avs}%
  \BibitemOpen
  \bibfield  {author} {\bibinfo {author} {\bibfnamefont {R.}~\bibnamefont
  {Egger}}\ and\ \bibinfo {author} {\bibfnamefont {H.}~\bibnamefont
  {Grabert}},\ }\href {\doibase 10.1103/PhysRevB.58.10761} {\bibfield
  {journal} {\bibinfo  {journal} {Phys. Rev. B}\ }\textbf {\bibinfo {volume}
  {58}},\ \bibinfo {pages} {10761} (\bibinfo {year} {1998})}\BibitemShut
  {NoStop}%
\bibitem [{\citenamefont {Recher}\ and\ \citenamefont
  {Loss}(2002)}]{recher2002-sct}%
  \BibitemOpen
  \bibfield  {author} {\bibinfo {author} {\bibfnamefont {P.}~\bibnamefont
  {Recher}}\ and\ \bibinfo {author} {\bibfnamefont {D.}~\bibnamefont {Loss}},\
  }\href {\doibase 10.1103/PhysRevB.65.165327} {\bibfield  {journal} {\bibinfo
  {journal} {Phys. Rev. B}\ }\textbf {\bibinfo {volume} {65}},\ \bibinfo
  {pages} {165327} (\bibinfo {year} {2002})}\BibitemShut {NoStop}%
\bibitem [{\citenamefont {V\"ayrynen}\ and\ \citenamefont
  {Ojanen}(2011)}]{vayrynen2011-ema}%
  \BibitemOpen
  \bibfield  {author} {\bibinfo {author} {\bibfnamefont {J.~I.}\ \bibnamefont
  {V\"ayrynen}}\ and\ \bibinfo {author} {\bibfnamefont {T.}~\bibnamefont
  {Ojanen}},\ }\href {\doibase 10.1103/PhysRevLett.106.076803} {\bibfield
  {journal} {\bibinfo  {journal} {Phys. Rev. Lett.}\ }\textbf {\bibinfo
  {volume} {106}},\ \bibinfo {pages} {076803} (\bibinfo {year}
  {2011})}\BibitemShut {NoStop}%
\bibitem [{\citenamefont {Str\"om}\ \emph {et~al.}(2010)\citenamefont
  {Str\"om}, \citenamefont {Johannesson},\ and\ \citenamefont
  {Japaridze}}]{strom2010-edi}%
  \BibitemOpen
  \bibfield  {author} {\bibinfo {author} {\bibfnamefont {A.}~\bibnamefont
  {Str\"om}}, \bibinfo {author} {\bibfnamefont {H.}~\bibnamefont
  {Johannesson}}, \ and\ \bibinfo {author} {\bibfnamefont {G.~I.}\ \bibnamefont
  {Japaridze}},\ }\href {\doibase 10.1103/PhysRevLett.104.256804} {\bibfield
  {journal} {\bibinfo  {journal} {Phys. Rev. Lett.}\ }\textbf {\bibinfo
  {volume} {104}},\ \bibinfo {pages} {256804} (\bibinfo {year}
  {2010})}\BibitemShut {NoStop}%
\bibitem [{\citenamefont {Golubov}\ \emph {et~al.}(2004)\citenamefont
  {Golubov}, \citenamefont {Kupriyanov},\ and\ \citenamefont
  {Il\char39{}ichev}}]{golubov2004-cri}%
  \BibitemOpen
  \bibfield  {author} {\bibinfo {author} {\bibfnamefont {A.~A.}\ \bibnamefont
  {Golubov}}, \bibinfo {author} {\bibfnamefont {M.~Y.}\ \bibnamefont
  {Kupriyanov}}, \ and\ \bibinfo {author} {\bibfnamefont {E.}~\bibnamefont
  {Il\char39{}ichev}},\ }\href {\doibase 10.1103/RevModPhys.76.411} {\bibfield
  {journal} {\bibinfo  {journal} {Rev. Mod. Phys.}\ }\textbf {\bibinfo {volume}
  {76}},\ \bibinfo {pages} {411} (\bibinfo {year} {2004})}\BibitemShut
  {NoStop}%
\bibitem [{\citenamefont {Della~Rocca}\ \emph {et~al.}(2007)\citenamefont
  {Della~Rocca}, \citenamefont {Chauvin}, \citenamefont {Huard}, \citenamefont
  {Pothier}, \citenamefont {Esteve},\ and\ \citenamefont
  {Urbina}}]{dellarocca2007-moc}%
  \BibitemOpen
  \bibfield  {author} {\bibinfo {author} {\bibfnamefont {M.~L.}\ \bibnamefont
  {Della~Rocca}}, \bibinfo {author} {\bibfnamefont {M.}~\bibnamefont
  {Chauvin}}, \bibinfo {author} {\bibfnamefont {B.}~\bibnamefont {Huard}},
  \bibinfo {author} {\bibfnamefont {H.}~\bibnamefont {Pothier}}, \bibinfo
  {author} {\bibfnamefont {D.}~\bibnamefont {Esteve}}, \ and\ \bibinfo {author}
  {\bibfnamefont {C.}~\bibnamefont {Urbina}},\ }\href {\doibase
  10.1103/PhysRevLett.99.127005} {\bibfield  {journal} {\bibinfo  {journal}
  {Phys. Rev. Lett.}\ }\textbf {\bibinfo {volume} {99}},\ \bibinfo {pages}
  {127005} (\bibinfo {year} {2007})}\BibitemShut {NoStop}%
\bibitem [{\citenamefont {Safi}\ and\ \citenamefont
  {Schulz}(1995)}]{safi1995-tts}%
  \BibitemOpen
  \bibfield  {author} {\bibinfo {author} {\bibfnamefont {I.}~\bibnamefont
  {Safi}}\ and\ \bibinfo {author} {\bibfnamefont {H.~J.}\ \bibnamefont
  {Schulz}},\ }in\ \href@noop {} {\emph {\bibinfo {booktitle} {Quantum
  Transport in Semiconductor Submicron Structures}}},\ \bibinfo {editor}
  {edited by\ \bibinfo {editor} {\bibfnamefont {B.}~\bibnamefont {Kramer}}}\
  (\bibinfo  {publisher} {Kluwer},\ \bibinfo {year} {1995})\ \bibinfo {note}
  {arXiv:cond-mat/9605014}\BibitemShut {NoStop}%
\bibitem [{\citenamefont {Dolcini}\ \emph {et~al.}(2005)\citenamefont
  {Dolcini}, \citenamefont {Trauzettel}, \citenamefont {Safi},\ and\
  \citenamefont {Grabert}}]{dolcini2005-tpo}%
  \BibitemOpen
  \bibfield  {author} {\bibinfo {author} {\bibfnamefont {F.}~\bibnamefont
  {Dolcini}}, \bibinfo {author} {\bibfnamefont {B.}~\bibnamefont {Trauzettel}},
  \bibinfo {author} {\bibfnamefont {I.}~\bibnamefont {Safi}}, \ and\ \bibinfo
  {author} {\bibfnamefont {H.}~\bibnamefont {Grabert}},\ }\href {\doibase
  10.1103/PhysRevB.71.165309} {\bibfield  {journal} {\bibinfo  {journal} {Phys.
  Rev. B}\ }\textbf {\bibinfo {volume} {71}},\ \bibinfo {pages} {165309}
  (\bibinfo {year} {2005})}\BibitemShut {NoStop}%
\end{thebibliography}
\end{document}